\documentclass[10pt,twocolumn,letterpaper]{article}
\usepackage[usenames, dvipsnames]{xcolor}
\usepackage{pifont}
\usepackage{usenix, times,color,graphicx,amsmath,array,amssymb,subcaption}
\usepackage{color}
\usepackage{colortbl}
\usepackage{multirow}
\usepackage{comment}
\usepackage{url}
\usepackage{amsmath}
\usepackage{wrapfig}
\usepackage{amsfonts}
\usepackage{amsthm}
\usepackage{listings}
\usepackage{xspace}
\usepackage{tightenum}
\usepackage{wrapfig}
\usepackage{adjustbox}
\usepackage{listings}
\usepackage{microtype}
\usepackage{tabulary}
\usepackage{upgreek}
\usepackage{floatflt}
\usepackage{mdframed} 
\usepackage{textcomp}
\usepackage[small,compact]{titlesec}
\usepackage{enumitem}
\usepackage[normalem]{ulem}
\usepackage[available,functional,reproduced]{usenixbadges}
\usepackage[numbers,sort&compress]{natbib} 
\usepackage[
  colorlinks=true,
  citecolor=red
]{hyperref}
\usepackage[T1]{fontenc}
\usepackage[utf8]{inputenc}
\usepackage{tikz}
\usepackage{mdframed}
\usepackage[ruled,linesnumbered]{algorithm2e}
\usepackage{colortbl}
\usepackage{booktabs}

\mdfdefinestyle{MyFrame}{%
    linecolor=black,
    outerlinewidth=2pt,
    innertopmargin=4pt,
    innerbottommargin=4pt,
    innerrightmargin=4pt,
    innerleftmargin=4pt,
        leftmargin = 4pt,
        rightmargin = 4pt,
    backgroundcolor=gray!40
}

\hypersetup{
  colorlinks,
  linkcolor={red!50!black},
  citecolor={blue!50!black},
  urlcolor={blue!80!black}
}

\newcommand{\squishlist}{
   \begin{list}{$\bullet$}
    { \setlength{\itemsep}{0pt}      \setlength{\parsep}{3pt}
      \setlength{\topsep}{3pt}       \setlength{\partopsep}{0pt}
      \setlength{\leftmargin}{1.0em} \setlength{\labelwidth}{1em}
      \setlength{\labelsep}{0.5em} } }
\newcommand{\squishend}{
    \end{list}  }
    
\newcommand{\tool}[0]{Atlas\xspace}

\definecolor{javared}{rgb}{0.6,0,0} 
\definecolor{javagreen}{rgb}{0.25,0.5,0.35} 
\definecolor{javapurple}{rgb}{0.5,0,0.35} 
\definecolor{javadocblue}{rgb}{0.25,0.35,0.75} 

\lstdefinestyle{mystyle}{
      language=C++,
        basicstyle=\scriptsize\ttfamily,
        keywordstyle=\color{javapurple}\bfseries,
        stringstyle=\color{javared},
        commentstyle=\color{javadocblue},
        morecomment=[s][\color{javadocblue}]{/**}{*/},
        numbers=left,
        breaklines=true,
        numberstyle=\tiny\color{black},
        stepnumber=1,
        numbersep=5pt,
        tabsize=2,
        showspaces=false,
        showstringspaces=false,
        morekeywords={foreach, uint64_t, uint16_t},
        classoffset=0,
        xleftmargin=1.8em,
        escapeinside={(*@}{@*)},
        moredelim=*[is][\color{red}]{[[[}{]]]},
        captionpos=b,
}
\newcommand{\upth}[0]{\textsuperscript{th}\xspace}

\newcommand{\codeIn}[1]{{\small\texttt{#1}}}

\newcommand{\MyPara}[1]{\vspace{.1em}\noindent\textbf{\textit{#1}}~}

\definecolor{ForestGreen}{RGB}{34,139,34}

\newcommand{\revise}[1] {{#1}}
\newcommand{\camera}[1] {{#1}}

\newcommand{\captionfonts}{\small}

\makeatletter  
\long\def\@makecaption#1#2{%
  \vskip\abovecaptionskip
  \sbox\@tempboxa{{\captionfonts #1: #2}}%
  \ifdim \wd\@tempboxa >\hsize
    {\captionfonts #1: #2\par}
  \else
    \hbox to\hsize{\hfil\box\@tempboxa\hfil}%
  \fi
  \vskip\belowcaptionskip}
\makeatother   

\newcommand{\squishlistree}{
   \begin{list}{$\bullet$}
    { \setlength{\itemsep}{0pt}      \setlength{\parsep}{0pt}
      \setlength{\topsep}{3pt}       \setlength{\partopsep}{0pt}
      \setlength{\leftmargin}{1em} \setlength{\labelwidth}{1em}
      \setlength{\labelsep}{0.5em} } }

\newcommand{\squishlisttwo}{
   \begin{list}{$\bullet$}
    { \setlength{\itemsep}{0pt}    \setlength{\parsep}{0pt}
      \setlength{\topsep}{0pt}     \setlength{\partopsep}{0pt}
      \setlength{\leftmargin}{2em} \setlength{\labelwidth}{1.5em}   
      \setlength{\labelsep}{0.5em} } }

\newcommand{\eg}{\hbox{\emph{e.g.}}\xspace}
\newcommand{\ie}{\hbox{\emph{i.e.}}\xspace}

\newcommand{\etc}{\hbox{\emph{etc.}}\xspace}

\makeatletter
\newcommand*{\circled}{\@ifstar\circledstar\circlednostar}
\makeatother
 
\newcommand*\circledstar[1]{%
   \tikz[baseline=(C.base)]
     \node[%
       fill=black!20,
       circle,
       minimum size=1em,
       text=black,
       font=\footnotesize,
       inner sep=0.3pt
     ](C) {#1};%
}
\newcommand*\circlednostar[1]{%
   \tikz[baseline=(C.base) - .6em]
     \node[%
       fill=black,
       text=white,
       circle,
       minimum size=.8em,
       font={\bf \footnotesize},
       inner sep=0.2pt
     ](C) {#1};%
}

\newcommand\blfootnote[1]{%
  \begingroup
  \renewcommand\thefootnote{}\footnote{#1}%
  \addtocounter{footnote}{-1}%
  \endgroup
}

\begin{document}
\pagenumbering{gobble}

\title{A Tale of Two Paths: Toward a Hybrid Data Plane \\for Efficient Far-Memory Applications\vspace{-1em}}
\author{\rm{Lei Chen}$^{\dag\ast}$\hspace{1.4em}Shi Liu$^{\psi\ast}$\hspace{1.4em}Chenxi Wang$^{\dag}$\hspace{1.4em}Haoran Ma$^{\psi}$\hspace{1.4em}Yifan Qiao$^{\psi}$\hspace{1.4em}Zhe Wang$^{\dag}$\\[.3em]\rm{Chenggang Wu}$^{\dag}$\hspace{1.4em} \rm{Youyou Lu}$^{\ddag}$\hspace{1.4em}Xiaobing Feng$^{\dag}$\hspace{1.4em}Huimin Cui$^{\dag}$\hspace{1.4em}Shan Lu$^{\theta}$\hspace{1.4em}Harry Xu$^{\psi}$
\\[0.5em]
University of Chinese Academy of Sciences$^{\dag}$\hspace{1em}UCLA$^{\psi}$\hspace{1em}Tsinghua University$^{\ddag}$\hspace{1em}Microsoft Research$^{\theta}$ 
}
\maketitle

\setcounter{page}{1}
\section*{Abstract}

With rapid advances in network hardware, 
far memory has gained a great deal of traction due to its ability to break the memory capacity wall.  
Existing far memory systems fall into one of two data paths: one that uses the kernel's paging system to transparently access far memory at the page granularity, and a second that bypasses the kernel, fetching data at the object granularity. While it is generally believed that object fetching outperforms paging due to its fine-grained access, it requires significantly more compute resources to run object-level LRU and eviction. 

We built \tool, a hybrid data plane enabled by a runtime-kernel co-design that simultaneously enables accesses via these two data paths to provide high efficiency for real-world applications. \tool uses \emph{always-on} profiling to continuously measure page locality. For workloads already with good locality, paging is used to fetch data, whereas for those without, object fetching is employed. Object fetching moves objects that are accessed close in time to contiguous local space, dynamically improving locality and making the execution increasingly amenable to paging, which is much more resource-efficient.
Our evaluation shows that \tool improves the throughput (\eg, by 1.5$\times$ and 3.2$\times$) and reduces the tail latency (\eg, by one and two orders of magnitude) \revise{when using remote memory}, compared with AIFM and Fastswap, the state-of-the-art techniques respectively in the two categories.

\blfootnote{$^\ast$ Contributed equally.}
\blfootnote{Corresponding authors: Chenxi Wang and Harry Xu.}
\section{Introduction}

Today's datacenters commonly suffer from low memory utilization~\cite{mem-harvesting-asplos22}; yet, datacenter applications are increasingly memory-constrained~\cite{itask-sosp15, borg-eurosys20,alibaba-cloud-bigdata17,lagar-cavilla-asplos19} due to their need to hold large datasets in memory for quick data analytics~\cite{spark,cassandra} or machine learning~\cite{tensorflow-osdi16, pytorch}. Thanks to the high bandwidth and low latency provided by modern network fabrics such as InfiniBand, far memory techniques~\cite{chenxi@jcst23,aifm@osdi2020,memliner-osdi22,infiniswap-nsdi17,fastswap-eurosys20,canvas-nsdi23} enable an abstraction of unlimited memory for applications by allowing them to use available memory on remote servers, thereby simultaneously improving application performance and datacenters' overall memory utilization.

Although techniques such as RDMA enable fast network accesses, each remote access is still at least an order of magnitude slower than a local access. As such, it is paramount to optimize the remote access data plane so that applications can benefit from increased memory capacity without suffering a significant performance hit. A major line of work for accessing remote memory is using the kernel's paging system, exemplified by techniques such as  InfiniSwap~\cite{infiniswap-nsdi17}, Fastswap~\cite{fastswap-eurosys20}, Canvas~\cite{canvas-nsdi23} and Hermit~\cite{hermit@nsdi23}.  These techniques allow applications to transparently access far memory at the \emph{page} granularity, using the kernel's swap system to swap pages in and out between local and remote memory.

While paging works well for applications that perform bulk data movement and exhibit clear (sequential or strided) access patterns, its coarse granularity incurs substantial \emph{I/O amplification} (\ie, pages loaded only contain a small amount of useful data) for applications that exhibit irregular (or random) access patterns, such as Memcached~\cite{memcached} and graph applications~\cite{graphone@fast19}. To reduce I/O amplification, a recent line of work exemplified by AIFM~\cite{aifm@osdi2020} and Kona~\cite{kona@asplos21} advocates to access data at a much finer (object) granularity using a user-space runtime system. Swapping objects, rather than pages, can significantly reduce the amount of useless data swapped, leading to higher efficiency. Furthermore, since objects are the data abstraction for developers to write programs, they carry semantics (\ie, user intention) that can be exposed to and used by the runtime to perform additional optimizations, such as data-structure-based prefetching. 

Fetching objects at runtime, however, comes at a cost.
A drawback that was often overlooked by existing works is that object fetching requires \emph{non-trivial compute resources} to profile object usage, identify patterns, and perform object-level LRU and eviction. For instance, running an object-level LRU algorithm is \textbf{one order of magnitude} more expensive than page-based LRU due to a huge number of objects to be processed and the lack of hardware support for tracking object accesses. This overhead is significantly more pronounced in real-world scenarios where CPUs are all busy with executing application threads\textemdash given a tight time budget, memory management threads cannot scan enough objects to make accurate LRU and eventually have to evict arbitrary objects. 

As a result, the right access mechanism is essentially the result of a tradeoff between program locality (\ie, how bad I/O amplification can be) and the amount of compute resources available (\ie, how many cores can be dedicated to object-level memory management tasks). For programs with poor locality, the overhead of object-level memory management can be offset from the large gains of reducing I/O amplification. On the other hand, for programs with good locality and insignificant I/O amplification, the overhead of object fetching stands out, especially in an environment where applications have taken all compute resources (see \S\ref{sec:motivation}). 

There is a recent line of compiler-based techniques (as exemplified by Mira~\cite{mira@sosp23}) that profile a program \emph{offline} to understand such a tradeoff, so that compiler can statically choose the mechanism for each data access when compiling the program. However, offline profiling hinges upon program input. For interactive applications such as Memcached, their input data comes from users and keeps changing, rendering a dry-run-based technique ineffective.

\MyPara{Major Insight.}  The main question we ask in this paper is: can we enable \emph{always-on} profiling for an application to identify its access patterns and dynamically switch between paging and object fetching to adapt to the observed patterns? This approach, if implemented efficiently, has two advantages over the state-of-the-art techniques.  First, its continuous profiling identifies patterns \emph{on-the-fly} for different computation stages or parallel threads accessing different data structures, even if the program input keeps changing. As a result, it can quickly change the access path to use a more efficient fetching mechanism. Second, for programs with irregular patterns, object fetching moves objects that are accessed close in time into contiguous memory space, dynamically improving locality as the program executes. This makes it possible for the execution to \emph{be increasingly amenable to paging}, which has higher resource efficiency (see \S\ref{sec:motivation}).

Although promising, realizing this insight requires overcoming three major challenges, as elaborated below:

The \emph{first challenge} is how to continuously and accurately profile an application with low overhead.
Kernel-based page-level profiling, though efficient, does not provide sufficient information with respect to fine-grained data locality. For example, if one single hot object on a page keeps getting accessed but none of other objects do, the kernel-based profiling would identify the page as a hot page although the page clearly possesses poor locality and its accesses should go through object fetching, not paging. 

To enable fine-grained profiling, \tool divides a page into a set of \emph{cards}, each of which is a unit for our locality measurement. We leverage the runtime (and in particular, a \emph{read barrier}) to compute a \emph{card access table (CAT)} (\S\ref{sec:memory_management})  for each page, which is a bitmap where each bit corresponds to a card (\ie, consecutive 16 bytes) on the page and a set bit represents that the card has been accessed since the page was allocated or last swapped in. 
A page with a high \emph{card access rate} (CAR, measured as the percentage of the set bits in its CAT) is deemed to possess good locality and should be accessed with paging, while a page with a low CAR has poor locality and should be accessed with object fetching.

The \emph{second challenge} is how to dynamically switch access mechanisms. 
\tool uses a read barrier at each smart pointer dereference. The barrier quickly checks a per-page \emph{path selector flag} (PSF) for the remote page to be accessed. \camera{Each PSF is a 1-bit flag, set to either \codeIn{runtime} or \codeIn{paging}. \codeIn{runtime} indicates that the runtime path should be used to fetch individual objects (like AIFM), while \codeIn{paging} means that the paging path is taken to fetch an entire page.} The PSF of a page is updated only when the page is evicted based upon the page's CAR\textemdash it is set to \camera{\codeIn{runtime}} if the page's CAR is low, indicating the page exhibits poor locality, and \camera{\codeIn{paging}} otherwise, indicating good locality.

Although \tool supports both object fetching and paging at \emph{ingress}, it evicts data only at the page granularity at \emph{egress}, to reduce the high overhead associated with object-level profiling and LRU. While evicting pages may introduce I/O amplification for workloads with poor locality, this impact is insignificant under \tool, because accesses in these workloads would likely go through the object fetching path, which improves locality by moving objects accessed close in time into contiguous local space. The enhanced locality effectively mitigates the negative impact of page-level eviction. 

To reduce fragmentation resulting from dead objects, \tool runs \emph{concurrent evacuation} tasks that periodically move live objects into contiguous memory space. During each evacuation, \tool groups recently-accessed objects into contiguous pages to further improve data locality.

The \emph{third challenge} is how to synchronize the two access paths. 
Since the kernel and the runtime are not coordinated (\eg, the kernel does not inform the runtime of the start or the completion of a page-fault handling), special care must be taken to prevent the two access paths from creating inconsistent data copies. In particular, correctness issues may arise from a set of ingress and egress events (\ie, object-in, page-in, and page-out) that occur simultaneously. \tool solves the problem with a synchronization protocol (see \S\ref{sec:runtime}), implemented with a combination of runtime and kernel support. 

\MyPara{Results.} 
We have evaluated \tool with a set of eight applications that cover a full range of memory access patterns: sequential, random, and mixed. Our results show that \tool enables these applications running on remote memory to achieve an overall of 1.5$\times$ and 3.2$\times$ throughput improvement, compared with AIFM~\cite{aifm@osdi2020} and Fastswap~\cite{fastswap-eurosys20}, respectively. 
\tool reduces the tail latency by one and two orders of magnitude when compared with AIFM and Fastswap. \tool is available at \href{https://github.com/wangchenxi7/Atlas}{https://github.com/wangchenxi7/Atlas}.

\section{Background on Object Fetching\label{sec:background}}

Object fetching is motivated by two observations on the inefficiencies of paging. First, fetching data at the page granularity often leads to I/O amplification~\cite{kona@asplos21}.
Second, managing data in the kernel space is agnostic to program semantics, resulting in missed optimization opportunities~\cite{aifm@osdi2020,memliner-osdi22,panthera@PLDI}. As such, work has been proposed to manage data with a language runtime at a finer-grained object (or cache-line) granularity~\cite{kona@asplos21,chenxi@jcst23,semeru@osdi2020,memliner-osdi22,aifm@osdi2020,mako@PLDI}. Unlike paging, the runtime can only manage objects in user space, which results in two consequences: (1) the runtime must change the virtual address of an object when moving it and hence must change all its pointers; and (2) the runtime must maintain all metadata itself (\eg, LRU), which used to be maintained by the kernel.
Here we focus our discussion on AIFM~\cite{aifm@osdi2020}. 
AIFM proposes two abstractions for developers to manage remote memory: \emph{remoteable pointer} and \emph{dereference scope}.

\MyPara{Remoteable pointer.} AIFM extends the \emph{smart pointer} abstraction of C++ to implement remoteable pointers (\codeIn{RemPtr}) for remote data management. There are two types of \codeIn{RemPtr}: 64-bit unique remoteable pointers (similar to \codeIn{std::unique\_ptr}) and 128-bit shared remoteable pointers (similar to \codeIn{std::shared\_ptr}). 
Developers need to explicitly declare data as remote type and manage them via the \codeIn{RemPtr}. For example, each unique \codeIn{RemPtr} has 64 bits\textemdash the lower 47 bits are used as the virtual address of the data, and the upper 17 bits are used to record metadata, such as dirty (D), present (P), hot (H), evacuated (E), \etc When accessing data via a \codeIn{RemPtr}, AIFM checks the metadata of the \codeIn{RemPtr} to detect its status, \eg, checking the P bit to see if the object is in local memory.
Next, AIFM masks the \codeIn{RemPtr} to obtain the actual virtual address.

\MyPara{Dereference scope.} Each smart pointer dereference and subsequent raw pointer accesses must be enclosed by a dereference scope, which works as an \emph{evacuation fence} to guarantee correctness. AIFM performs periodical \emph{concurrent object evacuation} that swaps out cold objects to remote memory and compacts local memory to improve data locality. 
It is challenging to move objects when they are being used by other threads since moving objects requires updating all their pointers. Smart pointers solve this problem because these pointers can be recorded in object headers and updated after moves are conducted. However, an application may read raw pointers from smart pointers and store them in registers or on the stack, which cannot be updated by the runtime.

To guarantee correctness for pointer updating, AIFM requires developers to explicitly declare dereference scopes for each object, which define where raw pointers of the object may exist. Evacuation of the object never happens concurrently with the execution of any of its dereference scopes that started before the evacuation decision. A dereference scope serves as a synchronization mechanism between an event that moves the object and another that uses it.

\section{Motivation \label{sec:motivation}}

We now motivate the necessity of a hybrid data plane. We first demonstrate the diverse memory access patterns of real-world cloud applications and explain the underlying reasons. Next, we compare fetching performance between using a runtime and the kernel's paging system.  
For the runtime approach, we re-implemented applications with AIFM~\cite{aifm@osdi2020}. For paging, we used Fastswap~\cite{fastswap-eurosys20}. Finally, we discuss the opportunities provided by \emph{dynamic} path switching.

\begin{figure*}[t]
    \centering
     \begin{tabular}{cccc}
    \includegraphics[scale=0.24]{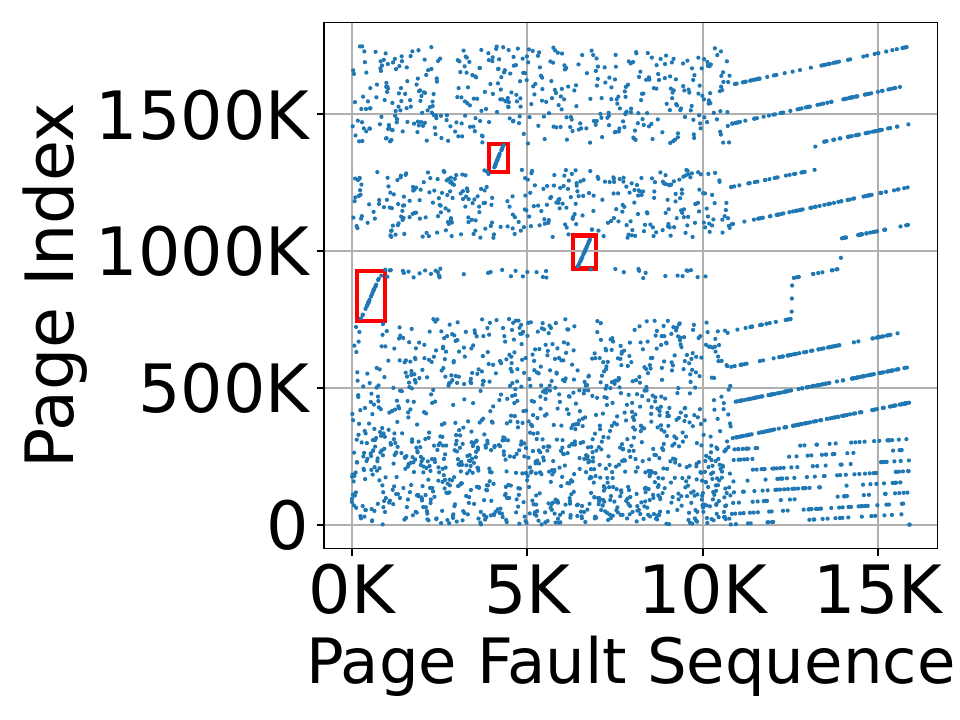} &
    \includegraphics[scale=0.23]{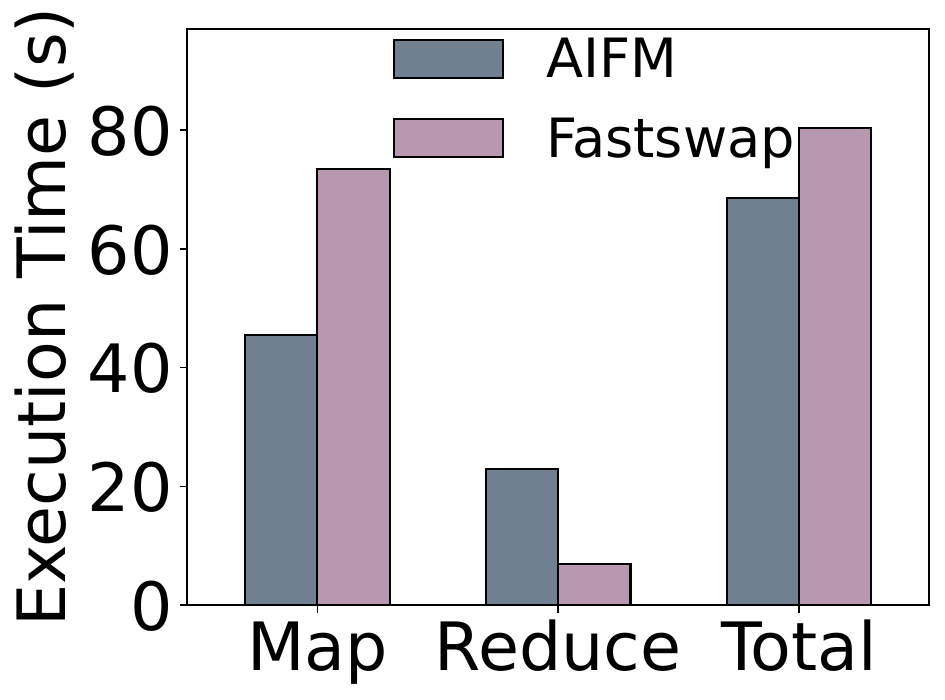} &
    \includegraphics[scale=0.18]{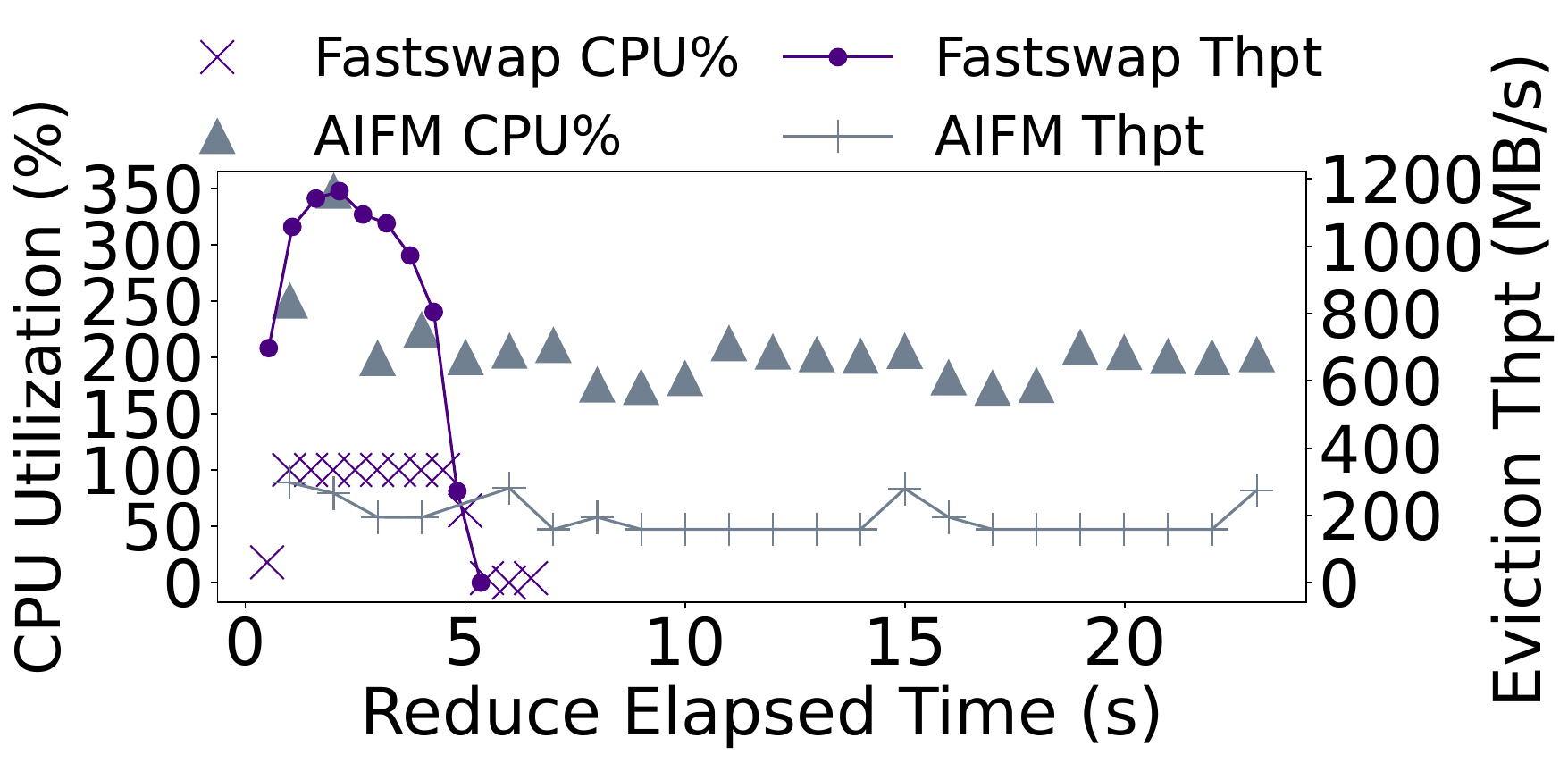} &
    \includegraphics[scale=0.24]{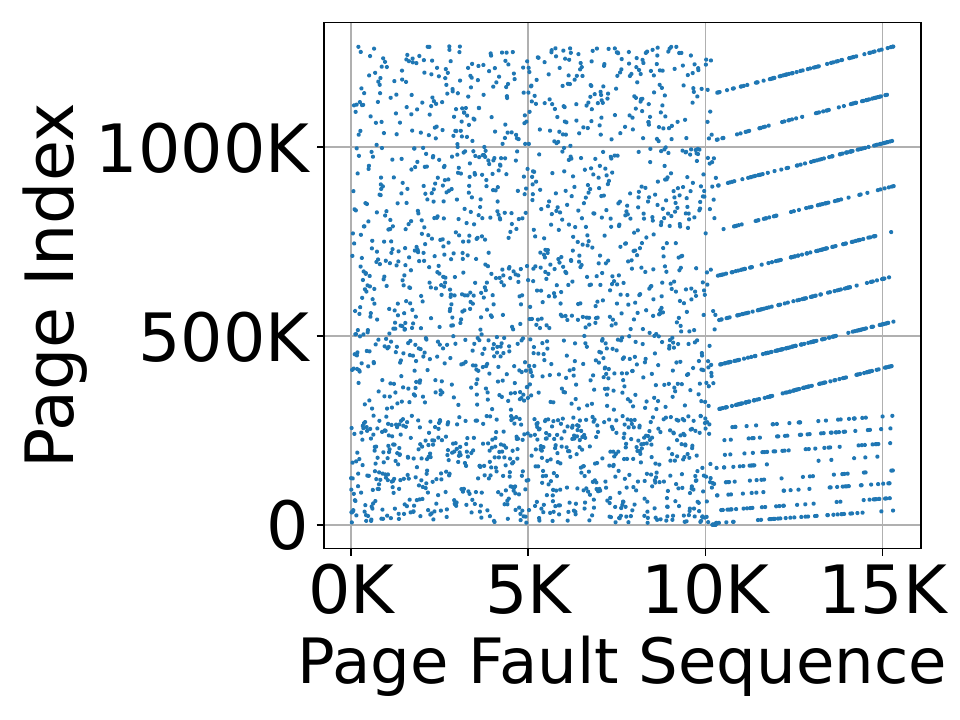} \\
    (a) Page Fault Trace 1  & (b) Throughput & (c) Memory \camera{Eviction} & (d) Page Fault Trace 2
    \end{tabular}
    \vspace{-1em}
    \caption{Statistics of Metis PageViewCount (MPVC): (a) access patterns, (b) performance comparisons between AIFM and Fastswap, (c) comparisons of \camera{eviction} throughput (dotted lines) and CPU usage (crosses and triangles) between AIFM and Fastswap, and (d) access patterns when input is changed to Wikipedia Italian~\cite{wikipedia-ds}. For these experiments, 25\% of the working set resides in the compute server's local memory. Sequential accesses (due to skewness) in the Map phase are highlighted in red boxes in (a), while in (d) such patterns do not exist. 
    \label{fig:motivation}}
    \vspace{-1.2em}
\end{figure*}

\MyPara{Diverse memory accesses.}
Real-world applications exhibit complicated memory access patterns, which are a combination of multiple primitive patterns such as sequential, strided, skewed, and random. Access patterns depend on at least two factors: (1) the computation model and (2) the data model. Next we elaborate on these factors:

On one hand, many applications are phase-changing and each phase follows a distinct computation model. On the other hand, the same phase may exhibit varied access patterns when processing different data structures.

An example is data-processing applications~\cite{hadoop-scheduling, spark} that implement MapReduce. We experiment with Metis~\cite{metis}, a MapReduce framework optimized for multicore architectures, with a Page View Count (PVC) program~\cite{mars@pact08,phoenix@hpca} and report its page fault sequence in Figure~\ref{fig:motivation}(a).  Since PVC is executed with 8 cores, we launch 8 threads for each (Map or Reduce) phase to exploit data parallelism. During the Map phase, each thread loads chunks of input data from the disk and initializes loaded website URLs and users as memory data. Next, PVC shuffles URLs into different buckets of a hash table based on their hash values.
The Reduce phase scans each entry to count each URL's users.

The left/right part of Figure~\ref{fig:motivation}(a) illustrates the page fault sequence of the Map/Reduce phase. 
The Map phase (left) inserts URLs into the hash table, and accesses there are mostly random. However, given that the dataset used to run this program is \emph{skewed}, there are several ranges of sequential accesses in the Map phase, as highlighted in the boxes (\ie, certain hash buckets are much larger than others and hence traversing these buckets exhibits sequential patterns). During the Reduce phase (right), each task that aggregates users of URLs scans entries in a bucket sequentially, resulting in a clear sequential access pattern, as shown in the second (right) half of Figure~\ref{fig:motivation}(a).

\MyPara{Granularity-performance tradeoff.}
Object fetching minimizes I/O amplification by fetching fine-grained objects~\cite{aifm@osdi2020,pberry@hotos}. However, compared to paging, object fetching does not always show clear benefits\textemdash for workloads with good locality, data on the same pages are accessed close in time and the kernel can already effectively and accurately prefetch data. When benefits are insignificant, the overhead for object-level memory management stands out. To compare fetching efficiency between the runtime and the kernel, we run the Metis PVC benchmark on AIFM and Fastswap, respectively. Figure~\ref{fig:motivation}(b) reports their performance comparisons.

Since a MapReduce program has clear phases, we broke down the execution time into Map and Reduce. AIFM outperforms Fastswap by 1.6$\times$ in the Map phase due to object fetching\textemdash  most remote accesses in Map are random as words are inserted into different buckets of the hash map. On the contrary, AIFM underperforms Fastswap by 3.3$\times$ in the Reduce phase, which exhibits clear sequential patterns.

\MyPara{Object eviction cost.} 
The main reason why object fetching underperforms paging for programs with good locality is the high cost associated with profiling objects and maintaining object-based LRU for eviction. For example, eviction must be done quickly as it blocks further memory allocations~\cite{hermit@nsdi23}. As a result, AIFM constantly maintains dozens of profiling/eviction threads to track the hotness of (billions of) objects and evict cold objects. However, if these threads cannot obtain enough CPU resources from the application, they end up scanning only a small percentage of objects before time runs out and then evict objects with limited hotness information, resulting in data thrashing (\ie,  hot objects get swapped out and quickly swapped back in). 

Figure~\ref{fig:motivation}(c) compares the eviction throughput and CPU utilization for eviction of AIFM and Fastswap during the Reduce phase. AIFM continuously performs object-level hotness tracking and eviction with around 200\% (up to 350\%) CPU usage in the entire Reduce phase. On the contrary, Fastswap finishes most of the page eviction task within the first five seconds and consumes no more than 100\% CPU resources during the eviction.
Overall, Fastswap consumes \emph{an order of magnitude} less compute (cycles) than AIFM for evictio over the Reduce phase. 
Even with significantly fewer CPU resources, Fastswap's eviction throughput is still $\thicksim$\textbf{5$\times$} higher than that of AIFM, due to the low memory management cost associated with paging.

\MyPara{Necessity of online profiling and path switching.}
Offline profiling techniques~\cite{mira@sosp23,weiwei@pact15,selective-page-migration,data-analysis-for-smp@pact10,llvm-analysis@cgo04} were proposed to analyze program semantics and data accesses. However, these techniques are ineffective in identifying the optimal solution for a real-world application for two major reasons. 

On the one hand, even if the application's computation phases may be analyzed by an offline profiling technique, its access patterns can change dramatically in response to \emph{inputs}. As Figure~\ref{fig:motivation}(d) demonstrates, when fed with a different dataset (which does not exhibit skewness), the program's access patterns change significantly\textemdash \eg, due to the lack of skewness, the Map phase no longer exhibits sequential patterns. In fact, for any interactive applications including Memcached~\cite{memcached}, DataFrame~\cite{dataframe}, or streaming data systems~\cite{graphone@fast19,aspen@pldi19,KickStarter@asplos17}, their behaviors and access patterns vary significantly with different user requests and workloads.

On the other hand, as discussed earlier, object fetching consumes extensive CPU resources. This may be acceptable when CPU resources are not fully saturated but becomes problematic as soon as all CPU cores are occupied (\eg, another tenant starts using the server). Clearly, offline profiling is not able to predict such environmental changes. 

These issues necessitate a dynamic technique that can continuously profile program executions and perform runtime data path switching as new behaviors and/or environmental changes are detected. Our main objective is to use object fetching to minimize I/O amplification and enhance locality, paving the way for subsequent accesses to operate on data with established locality and thus benefit from paging that is considerably more resource efficient.

\section{\tool Design and Implementation \label{sec:design}}
This section presents \tool's design. Like AIFM, \tool requires programs to use smart pointers (\ie, to implement barriers) and declare dereference scopes for objects (inspired by C++ weak
pointers~\cite{weakpointer} and Folly RCU guards~\cite{rcu}). Objects are managed by \tool's hybrid data plane.
\tool can also take the same user-defined programming/offloading hints and object-level prefetching logic as required by AIFM. \tool uses such hints in the object fetching path.

\subsection{Overview~\label{sec:overview}}

Inspired by the design of the Java heap~\cite{jvm}, \tool divides a page into \emph{cards} to enable fine-grained profiling for accesses. 
For each page, \tool builds a card access table (CAT), which is a bitmap where each set bit represents a card that has been accessed since the page was allocated or last swapped in. CATs for contiguous pages are allocated contiguously in a separate memory space. This design enables not only fine-grained access profiling, but also simple mapping from a virtual address to its CAT entry\textemdash this can be done with efficient bit-wise operations on the address. Each card represents 16 consecutive bytes, which provides a fine enough granularity as most objects are at least 16 bytes in our workloads.

\tool maintains \camera{a 1-bit} \emph{path selector flag} (PSF) for each page, which works as an indicator of the data path for data access on the page.  A \camera{\codeIn{runtime}} value indicates that data should be retrieved by the runtime at the object granularity (\ie, runtime path). A \camera{\codeIn{paging}} value indicates that data should be paged in by the kernel (\ie, paging path). \tool updates the PSF of each page to \camera{\codeIn{runtime or paging}} at the moment the page is swapped out if its CAR (\ie, the percentage of the set bits among all bits in a CAT) goes below or above a threshold (\ie, 80\% used in our evaluation, see \S\ref{sec:drill-down}).

\MyPara{Ingress.} \tool uses a \emph{read barrier} that executes at each \emph{smart pointer dereference}. The barrier first checks whether the accessed data is remote.  In AIFM, this is done by using a bit in each pointer to encode the location of the referenced object\textemdash these pointers are updated once the objects they point to are swapped in or out. \tool, however, cannot adopt this approach due to the use of the hybrid data plane\textemdash when data is \emph{paged} out, \tool cannot update any pointers. To solve the problem without incurring the cost of checking with the kernel at every read, \tool leverages \emph{hardware transaction memory} and, in particular, Intel's TSX~\cite{intel-sdm-tsx}, to run a quick check\textemdash
\tool accesses the address in a hardware transaction, which aborts if the address is not on a mapped page. 

Upon an abort, the barrier reads the PSF for the page to be accessed and determines which path (runtime \emph{vs.} paging) the access should take. If the runtime path is taken, the object is moved (\ie, address changed) to a local page on the compute server and its pointers are updated; otherwise, the page containing the object is swapped in as a whole and the address of the object remains the same (without requiring pointer updating).

\MyPara{Egress.} Given that the majority of the object-fetching overhead comes from the need to find and evict cold objects, \tool utilizes a single path, \ie, paging, to swap out data.
This approach achieves a sweet spot in balancing overhead and benefits\textemdash on one hand, it significantly reduces the compute resource usage for object fetching because of the elimination of maintaining an object-level LRU; on the other hand, given that object fetching gradually improves locality (by moving together objects accessed closely in time), the amount of useless data in each swap-out (and thus the I/O amplification) is reduced progressively during execution. 

Another reason to not evict objects individually is that it can potentially \emph{hurt locality}\textemdash after objects are fetched in, those that were scattered in remote memory but accessed together were moved into contiguous local space; however, these objects may not be evicted at the same time; evicting them individually would make them go to unrelated locations in remote memory, disrupting established locality.

\MyPara{Synchronization.} Allowing the two paths to co-exist in harmony requires overcoming the following three synchronization challenges: (1) \emph{ingress synchronization} between object-in and page-in, (2) \emph{egress synchronization} between object-in and page-out, and (3) \emph{move synchronization} between object-in and evacuation. AIFM already solves the third problem with the declaration of dereference scopes, while the other two are unique challenges that we target in \tool.

\subsection{Synchronization of the Two Paths~\label{sec:runtime}} 
\tool builds its object fetching path upon the same two abstractions used by AIFM: the smart pointer (which is an extension of C++ smart pointer) and the dereference scope. This section elaborates on the synchronization mechanism between the object fetching path and the paging path.

\begin{figure}[h]
\vspace{-1em}
\small
\begin{lstlisting}[language = c]
  class AtlasUniquePtr<T>{
    struct AtlasMetadata{
      unsigned long is_moving : 1;
      unsigned long access : 1;
      unsigned long reserve : 2;
      unsigned long offload : 1;
      unsigned long size : 12;
      unsigned long addr : 47;
    } metadata; // 64 bits
    AtlasUniquePtr(T* obj);
    T* get_raw();
  }
\end{lstlisting}
\vspace{-1.5em}
\caption{\revise{\tool unique pointer metadata.}\label{fig:atlas-unique-ptr}}
\vspace{-.5em}
\end{figure}

\MyPara{Pointer Metadata.} 
Before discussing our barrier logic, we first present the format of \tool pointers, which are built on C++ smart pointers.  \tool uses two types of smart pointers: unique pointers (similar to \codeIn{std::unique\_ptr}) and shared pointers (similar to \codeIn{std::shared\_ptr}). Figure~\ref{fig:atlas-unique-ptr} shows the layout of an \tool unique pointer. These fields are added for the purpose of synchronization and pointer updating. 

Each such pointer has 64-bit metadata, in which 47 bits (\codeIn{addr}) store the object's raw pointer, 12 bits (\codeIn{size}) record its size, 1 bit (\codeIn{access}) represents whether the object has been accessed since the last evacuation (which will be used by the evacuator to group recently accessed objects, see \S\ref{sec:memory_management}), 1 bit (\codeIn{offload}) indicates whether a function is being invoked on the object on the remote side, and 1 bit (\codeIn{is\_moving}) indicates whether the object is being moved (\eg, due to evacuation); this bit will be used for synchronization between two threads trying to move the same object. The remaining 2 bits (\codeIn{reserve}) are reserved for future use. Note that 12 bits can represent a size up to 4KB. Objects larger than that are placed in the huge-object space of the heap for which paging is the only option. \codeIn{get\_raw} retrieves the raw pointer from a smart pointer.

A shared pointer allows aliasing. \tool treats the first shared pointer of an object as the main pointer. A shared pointer's layout is similar to a unique pointer, except that it has an additional 8 bytes to chain all pointers\textemdash when the main pointer is being released, \tool follows the chain to select a new main pointer. If an object is referenced by shared pointers, \tool needs to update all of them (by following the chain).

Developers need to explicitly declare data types with smart pointers. Developers can access data with raw (regular C++) pointers by first retrieving such raw pointers from smart pointers. However, this can only be done within an explicitly declared dereference scope. Figure~\ref{fig:unique-ptr-example} illustrates an example of retrieving and manipulating data from \tool smart pointers, confined by a dereference scope. As discussed in \S\ref{sec:background}, dereference scopes synchronize with object migration tasks\textemdash once raw pointers are retrieved and actively used, their objects are not allowed to move, and vice versa. \tool executes a \codeIn{pre\_scope\_barrier} and a \codeIn{post\_scope\_barrier} at the beginning and the end of the dereference scope, respectively. 

\begin{figure}[h]
\vspace{-1em}
\small
\begin{lstlisting}[language = C]
  deref_scope (smart_ptr) {
    pre_scope_barier(smart_ptr); // Algorithm 1
    Data * object = smart_ptr.get_raw(); 
    /* Operations using the object */
    ... 
    post_scope_barrier(smart_ptr); // Algorithm 2
  } 
\end{lstlisting}
\vspace{-1em}
\caption{Dereferencing an \tool unique pointer in a deref scope. 
\label{fig:unique-ptr-example}}
\vspace{-1em}
\end{figure}

\MyPara{Synchronization invariants.} We present a set of high-level invariants that \tool maintains to solve the three synchronization problems: (1) preventing an object from being fetched from the two paths simultaneously (object-in \emph{vs.} page-in), (2) preventing pages containing objects that were just runtime-fetched from being immediately swapped out (object-in \emph{vs.} page-out), and (3) preventing an object from being simultaneously runtime-fetched and moved by the evacuator (object-in \emph{vs.} evacuation).

\emph{Invariant \#1: Object-in vs. page-in.} At any moment, all data on the same page must go through the same access path as guided by the page's PSF. In other words, \tool prohibits scenarios where certain requests are served by paging while others are served by the runtime for the same page. Given that \tool changes PSF only at page-out (as opposed to setting it while the page is in local memory), such scenarios can never occur and this invariant is guaranteed by design. 

Note that there is no issue if two threads fetch the same page from the paging path\textemdash the kernel's swap system guarantees only one page can be mapped. Fetching the same object from two threads with the runtime path is not a concern either: it is a solved problem in the literature of moving garbage collectors~\cite{mako@PLDI} where pointer updating is used as a synchronization point and only one object is retained.

\emph{Invariant \#2: Object-in vs. page-out.} Since swap-out events can occur at any time with the runtime path uninformed, \tool enforces that pages containing objects whose dereference scopes are actively executed cannot be swapped out. This is because if such pages are swapped out before their dereference scopes finish, these objects may be fetched back in immediately from the runtime path, requiring pointer updating. Pointer updating cannot be done when the raw pointers of these objects are active on the stack. As a result, these pages cannot be swapped out until none of their objects are executing their dereference scopes.

\tool achieves this by maintaining a per-page \emph{deref count}, which is incremented when any object on the page enters a dereference scope and decremented when the scope finishes. Any page with a non-zero deref count is skipped when the kernel looks for swap-out victims. Note that this does not create much impact on performance because the pages whose objects are actively used are usually hot pages and unlikely to be selected as swap-out victims anyway.

One issue that may arise from this protection is a potential live lock on the object-fetching path: either an ill-defined large dereference scope or many active dereference scopes in a parallel application may potentially lead to too much data getting pinned in local memory, which may result in out-of-memory errors.  To tackle this issue, \tool monitors the pinned data and forces the flipping of their containing pages' PSFs (to use paging) upon memory pressure. Once these pages are swapped out, they will be paged in\textemdash this solves the problem as page-in does not need pointer updating.

\begin{algorithm}[t]
 \scriptsize
 \caption{\tool Pre-Scope Barrier (Simplified).\label{algo:pre_barrier}}
   \SetKwData{Left}{left}\SetKwData{This}{this}\SetKwData{Up}{up}
   \SetKwData{NewMeta}{new\_value}\SetKwData{Meta}{metadata}\SetKwData{Addr}{addr}
   \SetKwData{OldMeta}{old\_value}
   \SetKwData{FaultFlag}{pff}
   \SetKwData{MoveFlag}{is\_moving}
   \SetKwData{Temp}{temp}
   \SetKwData{R}{r}
   \SetKwData{Ref}{ref}
   \SetKwData{OSize}{size}
   \SetKwData{NewAddr}{new\_addr}
   \SetKwData{Thres}{THRESHOLD}
   \SetKwData{NotS}{not}\SetKwData{AndS}{and}
   \SetKwData{OrS}{or}
   \SetKwData{Eva}{is\_moving}
   \SetKwFunction{SetFlag}{set\_flag}
   \SetKwFunction{ClearFlag}{clear\_flag}
   \SetKwFunction{SetEva}{set\_move}
   \SetKwFunction{ClrEva}{clear\_move}
   \SetKwFunction{GetAddr}{get\_addr}
   \SetKwFunction{GetSize}{get\_size}
   \SetKwFunction{SetAddr}{set\_addr}
   \SetKwFunction{TSX}{tsx\_check\_local}
   \SetKwFunction{PSF}{take\_runtime\_path}
   \SetKwFunction{Inc}{inc\_ref}
   \SetKwFunction{XCHG}{compare\_and\_swap}
   \SetKwFunction{FindPage}{find\_page\_meta}
   \SetKwFunction{AtomInc}{atom\_inc}
   \SetKwFunction{AtomDec}{atom\_dec}
    \SetKwFunction{CompAddr}{find\_addr}
      \SetKwFunction{ObjIn}{alloc\_copy\_update}
   \SetKwInOut{Input}{input}\SetKwInOut{Output}{output}
     \SetKwFunction{FindVR}{find\_vr}
      \SetKwFunction{DecIFR}{atom\_dec\_ifr}
   \BlankLine
\tcc*[h]{derefcnt > 0 precludes the page's swap-out}\\
           \AtomInc{\FindPage{\Addr}.derefcnt} \\  \label{l:page-count-inc}

       \If(\tcc*[f]{Remote object}){ \NotS \TSX{\Addr}  \label{l:tsx}}{
         
         \If(\tcc*[f]{Runtime path}){\PSF{\Addr} \label{l:psf}}{

         \NewAddr $\leftarrow$ \CompAddr{\Addr, \This.size}\\ \label{l:comp-addr}
                  \tcc*[h]{Inc/dec the new/old page's derefcnt}\\ 
                   \AtomInc{\FindPage{\NewAddr}.derefcnt} \\ \label{l:inc-new-page-count}
        \AtomDec{\FindPage{\Addr}.derefcnt} \\ \label{l:dec-page-count}
         \ObjIn{\Addr, \NewAddr, \This.size} \\ \label{l:fetch}
         \This.\Meta.addr $\leftarrow$ \NewAddr \\ \label{l:setaddr}
        
         \Addr $\leftarrow$ \NewAddr \label{l:end-runtime}
         } 
       \Else(\tcc*[f]{Paging path}){                    
            * (char*) \Addr \label{l:deref}
       
        }
       
       }
    
 \end{algorithm}

\begin{algorithm}[t]
 \scriptsize
 \caption{\tool Post-Scope Barrier.\label{algo:post_barrier}}
   \SetKwData{Left}{left}\SetKwData{This}{this}
   \SetKwFunction{FindPage}{find\_page\_meta}
   \SetKwFunction{AtomDec}{atom\_dec}
   \SetKwFunction{Store}{store}
   \SetKwInOut{Input}{input}\SetKwInOut{Output}{output}
   \BlankLine
       \AtomDec{\FindPage{\This.addr}.derefcnt} \\
 \end{algorithm}

\emph{Invariant \#3: Dereference scope vs. evacuation.} Evacuation threads may move an object while another thread is executing the object's dereference scope. This must not occur because evacuation requires pointer updating, which cannot be done when a dereference scope is being executed (and raw pointers are used). To this end, \tool uses the page's deref count to synchronize between evacuation threads and dereference scopes. A non-zero dereference count prevents the page from being evacuated. 

\revise{Compared to AIFM, \tool employs a slightly different definition of dereference scope. AIFM chose to decouple dereference scopes from the barrier\textemdash it allows one dereference scope to cover multiple smart pointer dereferences, serving as a coarse-grained fence between application threads and the evacuator. 
On the contrary, \tool employs \emph{fine-grained} dereference scopes, each of which is associated with one single smart pointer dereference. This choice was made based on our observation of frequent evacuations; using coarse-grained dereference scopes would require constant synchronizations between application and evacuation threads, 
leading to performance and latency impact. 
Fine-grained dereference scopes not only reduce the degree of blocking but also help alleviate potential live locks. Although a finer granularity increases barrier overhead, this overhead is often amortized by a large number of raw pointer accesses and computation within each scope. A detailed overhead analysis can be found in \S\ref{sec:eval:tput} and \S\ref{sec:drill-down}.}

With the invariants discussed above, we proceed to presenting our barrier logic, which is shown in Algorithm~\ref{algo:pre_barrier} and Algorithm~\ref{algo:post_barrier}. As illustrated in Figure~\ref{fig:unique-ptr-example}, \tool executes Algorithm~\ref{algo:pre_barrier} and Algorithm~\ref{algo:post_barrier} at the beginning and the end of a dereference scope, respectively. 

\MyPara{Pre-scope barrier.} 
\tool first atomically increments the deref count for the page containing the object (Line~\ref{l:page-count-inc}). 
This indicates that the page has an object whose dereference scope is being executed, preventing the paging system from swapping out the page (\ie, Invariant \#2). This step must be done before the barrier starts to guarantee that (1) if the page is local, it cannot be swapped out from this point on, or (2) if the object is remote, once it is fetched in, its containing page cannot be swapped out.

\tool uses Intel's TSX~\cite{intel-sdm-xbegin}  
to efficiently check if the address \codeIn{addr} is local.  
\tool starts an RTM transaction, which contains nothing but a dereference of the object.
If the object's containing page is unmapped, the RTM transaction will abort with a special status captured by \tool, which verifies the status by checking with the kernel. This hardware-based check is $\thicksim$14$\times$ faster than a purely software-based approach that relies on a system call that walks the page table and checks whether the page is local based on its PTE. 
A \codeIn{true} value (\ie, the object is local) returned by TSX directs the execution to exit the barrier immediately. Otherwise, \tool checks the PSF corresponding to the address (Line~\ref{l:psf}) to decide whether this access should take the runtime (Lines \ref{l:comp-addr}-\ref{l:end-runtime}) or the paging path (Line~\ref{l:deref}). 

Using TSX to check object location may introduce false positives\textemdash a transaction may abort even if data is local. Since such cases are rare (\eg, less than 1/10000 in our experiments), \tool takes an optimistic approach to handle them. Upon a TSX abort, \tool sends an RDMA read to access the remote object and simultaneously issues a page table walk to verify the object's location. If the verification fails (indicating the object is local), the fetched object is discarded. This approach introduces only a negligible overhead (\ie, a small number of unnecessary RDMA reads).

\MyPara{Runtime path.} \codeIn{take\_runtime\_path} in Algorithm~\ref{algo:pre_barrier} checks the PSF of the page corresponding to \codeIn{addr} and returns \codeIn{true} if the PSF is \codeIn{runtime}, indicating that object fetching should be performed. For ease of presentation,  Algorithm~\ref{algo:pre_barrier} is significantly simplified to \emph{not} show details of how to synchronize between threads to guarantee the absence of race condition when multiple threads fetching the same object. 
\tool first finds a new address to which the object will be moved (Line~\ref{l:comp-addr}). Since this address is on a new page, before moving the object, the deref count of the new page must be incremented (Line~\ref{l:inc-new-page-count}) to ensure that from this point on, the new page cannot be swapped out until the dereference scope finishes (\ie, Invariant \#2). The barrier also needs to decrement the deref count of the old page (Line~\ref{l:dec-page-count}) that was incremented earlier in Line \ref{l:page-count-inc}.

Next, \tool fetches the object by allocating a new object of the same size (using our log-structured allocator discussed in \S\ref{sec:memory_management}), copying the object's data into the new object, and updates its pointers (Line~\ref{l:fetch}). 
\tool subsequently changes the \codeIn{addr} field of the pointer to the new address (Line~\ref{l:setaddr}). Pointer updating is done by retrieving the object's pointer from its header and updating their addresses, in a way similar to how it is done in AIFM. If it is a shared pointer, all other pointers will be retrieved from the main one and updated accordingly. The object's \codeIn{is\_moving} field is used to synchronize between pointer updating events performed by multiple threads. The synchronization details are omitted for simplicity. 
After the object is moved to a local page, future accesses to the object will follow the PSF of the new page.

\MyPara{Paging path.} The paging path simply touches the object (Line~\ref{l:deref}) to ensure that the page fault handling is \emph{completed} after the execution passes this line.

\MyPara{Post-scope barrier.} 
The post-scope barrier has much simpler logic, as shown in Algorithm~\ref{algo:post_barrier}. All it needs to do is to atomically decrement the page's deref count, indicating the finishing of the dereference scope. When its deref count becomes zero, this page is subject to swap-out again (\ie, Invariant \#2).

\subsection{Memory Management\label{sec:memory_management}}

\tool's heap is composed of a \emph{normal-object} space, a \emph{huge-object} space, a \emph{metadata} space, and an \emph{offload} space.  \tool manages the normal-object space via a log-structured allocator~\cite{lsm@fast14, aifm@osdi2020} and maintains a background evacuator to reduce fragmentation by compacting live objects. \tool does not handle huge objects that cannot fit into a page, placing them into the huge-object space and delegating their management to the kernel directly since they are too large to benefit from object-level management.  Metadata such as CATs are accessed by both the runtime and paging system, and hence, it is shared between the user and kernel space. The offload space stores objects whose functions can be offloaded to the remote side. We will discuss it shortly.

\MyPara{Object allocation.} 
The log-structured allocator maintains thread-local allocation buffers (TLAB) to reduce the global lock contention during parallel object allocation. The TLAB is managed at the granularity of log segment which is aligned with a page to guaranteed that no object can go cross the page boundary. 
\tool allocates objects contiguously on the TLAB as prior research~\cite{weiwei@pact15, panthera@PLDI} shows that objects allocated close in time exhibit similar usage patterns. In doing so, objects with temporal proximity are naturally grouped into the same log segment (page), enhancing locality.

\MyPara{Metadata allocation.}
Metadata such as dereference counters and card tables is allocated in a dedicated metadata space.
\tool maintains a card table for each page to record the object access information. Each card table is a bitmap where each bit represents a consecutive range of 16 bytes. Our experiments show that the sizes of most objects are larger than 8 bytes, making 16 bytes a natural choice for the card size. 
Each card table is allocated and initialized during the allocation of a log segment. It is freed along with the log segment. The space needed by the card tables is 1/128 of the total memory. In summary, the space overhead is less than 2\%.

\MyPara{Object evacuation.}
The log-structure allocator~\cite{lsm@fast14} supports defragmentation via a copying-based evacuator, a technique widely used in modern garbage collectors~\cite{Shenandoah}. In \tool, we extend the evacuator to improve the temporal locality of pages by grouping hot objects into contiguous log segments (pages) during the evacuation. The evacuator runs concurrently with the application to reduce fragmentation. 

The evacuator periodically scans log segments and evacuates a log segment with a high garbage ratio by copying its live objects to a newly allocated target segment. As a result, the target segment is free of fragmentation, and the source log segment can be freed right away. When moving an object, the evacuator maintains its corresponding card table values, \ie, if the object was recently accessed on the source page, the evacuator marks its card bit on the target page during evacuation.  
Furthermore, \tool improves evacuation efficiency by prioritizing log segments in local memory and delaying the processing of remote log segments until they are accessed or the free space runs out~\cite{memliner-osdi22}.  

The \tool runtime tracks whether an object has been accessed since the last evacuation via the \codeIn{access} bit in the smart pointer (see Figure~\ref{fig:atlas-unique-ptr}). This bit is set by the read barrier when the object is dereferenced and cleared by the evacuator at the end of each evacuation. The evacuator segregates objects that have been accessed since the last evacuation into a set of contiguously allocated log segments. We found this approach to be particularly effective in improving temporal locality for real-world workloads with \emph{skewness} (\eg, 90\% of accesses hit 10\% objects).
The \codeIn{access} bit allows \tool to distinguish hot and cold objects in such workloads, leading to a substantial performance boost.
Note that this operation is significantly more efficient than maintaining an object-level LRU for eviction. As opposed to ranking objects based on hotness, \tool's \codeIn{access} bit simply serves as an evacuation location indicator. Its functionality is similar to CAT but used differently; CAT is read and cleared by the kernel at page eviction while the \codeIn{access} bit is read and cleared by the runtime at evacuation.

\MyPara{Computation offloading.}
As shown in many existing far-memory systems, such as Semeru~\cite{semeru@osdi2020}, Mako~\cite{mako@PLDI}, AIFM~\cite{aifm@osdi2020}, and Mira~\cite{mira@sosp23}, offloading memory-intensive operations to the remote side can effectively reduce the data movement overhead. A unique challenge for \tool is how to enable offloading when paging is used. Under paging, remote memory is managed as a swap partition of a set of swap slots. These slots are agnostic about the remote server's memory addresses. Pointer addresses contained in a page are with respect to the compute server while the page can reside at a completely different address on the remote server. This address mismatch precludes the correct execution of a function on an object directly on the remote server. 

To solve the problem, \tool uses an approach that is similar to Semeru~\cite{semeru@osdi2020}\textemdash we reserve a dedicated offload space in the heap. Developers need to explicitly define remoteable data structures and functions (which are similar to those in AIFM). Objects registered as \emph{remotable} are all allocated into this space. Pages in this space have guaranteed virtual address alignment between the compute and remote servers\textemdash we modify the paging system to ensure that a page at a virtual address A on the compute server is guaranteed to be still at address A on the remote server when evicted. \tool requires users to guarantee a remotable data structure cannot reference a non-remotable object. This property ensures address consistency when a function is called remotely.  

The offload space is an \emph{object-in, page-out} space, which allows objects to be fetched only through the runtime. This is due to the need to synchronize between the servers for safe remote execution. When a remote function is being invoked on an object, the \codeIn{offload} field in its smart pointer is used for synchronization\textemdash the runtime can not fetch the object until the remote function is finished (and the \codeIn{offload} bit is cleared). Remotable objects can only be fetched into the offload space to ensure the above-stated properties.

\section{Evaluation \label{sec:evaluation}}

\subsection{Setup and Methodology~\label{sec:setup}} 
We wrote 7,675 lines of C/C++ code to implement \tool's runtime library, and added support in the Linux kernel (version 5.14-rc5) for page management (\eg, path synchronization). 
We ran experiments with one compute server and one memory server connected by a 200 Gbps Infiniband switch. 
Each server has 2 Intel Xeon Gold 6342 CPUs (24 physical cores each), 256 GB of memory, and a 100 Gbps Mellanox ConnectX-5 InfiniBand adapter. All evaluated systems ran on Ubuntu 18.04. We configured the servers following common practice for low latency~\cite{shenango-nsdi19}, disabling Turbo Boost, CPU frequency scaling, and transparent huge pages.

\MyPara{Baselines.}
\tool was implemented based on Fastswap and AIFM. For the paging path, \tool uses unmodified Fastswap with added tasks of profiling and synchronization. For the runtime path, Atlas uses AIFM's ingress algorithm and paging at egress.  
For evaluation, we used AIFM~\cite{aifm@osdi2020} and Fastswap~\cite{fastswap-eurosys20} as our baselines for object fetching and paging, respectively. 
For Fastswap, we ran the original applications to avoid unnecessary runtime overhead. For AIFM, we used the performance-tuned versions of applications, where all optimizations were enabled including per-thread access pattern tracking, object hotness tracking, and non-temporal programming hints~\cite{aifm@osdi2020}.
We turned off offloading when evaluating throughput and latency, leaving its evaluation to \S\ref{sec:drill-down}.

\begin{table*}[t]
\centering
\begin{adjustbox}{max width=\textwidth}
\small

\begin{tabular}{c|c|c|c}

      \rowcolor[HTML]{EFEFEF} 
      \hline
      \textbf{Application}  & \textbf{Dataset} & \textbf{Size} & \textbf{Characteristics} \\ \hline

    {Memcached CacheLib ~\cite{memcached} (\textbf{MCD-CL})}  & Meta CacheLib~\cite{cachelib@osdi20} & 50M records &  Skewness with churn \\
    {Memcached Uniform (\textbf{MCD-U})}  & Synthetic, uniform distribution~\cite{cooper-ycsb-ds} & 50M records & Random access \\

    GraphOne PageRank~\cite{graphone@fast19} (\textbf{GPR})   & Twitter 2010~\cite{Kwak10www} & 1.5B Edges, 41.7M Vertices & Evolving graph \\

    Aspen TriangleCount~\cite{aspen@pldi19} (\textbf{ATC}) & Friendster~\cite{snap} & 1.8B Edges, 65.6M Vertices & Evolving graph \\

    Metis Word Count~\cite{metis} (\textbf{MWC}) &  The News Crawl Corpus~\cite{wc_input} &  5.1GB & Phase-changing \\
    
    Metis PageViewCount (\textbf{MPVC}) &   Wikipedia English~\cite{wikipedia-ds} & 15GB & Phase-changing with mixed patterns \\
    
    DataFrame~\cite{dataframe} (\textbf{DF}) & NYC Taxi~\cite{nyctaxi}  & 16 GB & Phase-changing with offloading\\

    Web Service~\cite{aifm@osdi2020} (\textbf{WS}) & Synthetic~\cite{aifm@osdi2020}   & 10GB hashmap, 16GB array & Mixed patterns with offloading \\

      \hline
             
\end{tabular}
\end{adjustbox}
\vspace{-1em}
\caption{\label{tab:real-apps} Applications used for our evaluation.}
\end{table*}

\MyPara{Workloads.}
As shown in Table~\ref{tab:real-apps}, we evaluated six real-world applications and two synthetic applications, including
Metis~\cite{metis}\textemdash an optimized MapReduce framework for multicore architectures, Aspen~\cite{aspen@pldi19}\textemdash a purely functional tree-based graph processing framework, GraphOne~\cite{graphone@fast19}\textemdash a data store for real-time analytics on evolving graphs,  as well as Memcached~\cite{memcached}\textemdash an in-memory key-value store. We ran Memcached with two different workloads: a real-world workload (MCD-CL) that comes from Meta's cache system CacheLib~\cite{cachelib@osdi20} and a synthetic workload (MCD-U) generated by YCSB~\cite{cooper-ycsb-ds} that follows a uniform distribution. We also employed two synthetic applications developed by AIFM's authors to compare \tool and AIFM. These applications include one batch application, DataFrame~\cite{dataframe}, and one latency-critical application, WebService. 

Covering a wide spectrum of domains and memory access patterns (\ie, sequential, random, skewed, and mixed patterns), these applications can be divided into four categories:

First, both Memcached workloads exhibit random access patterns, leading to significant I/O amplification under paging. The real-world workload MCD-CL has a high level of skewness with \emph{churn} behaviors. \emph{Churn} refers to the phenomenon that hot data in the working set changes rapidly over time. On the contrary, the synthetic workload MCD-U demonstrates completely random behaviors, with no skewness and hot data. As a result, MCD-CL is more amenable to \tool's dynamic locality improvement than MCD-U. 

Second, GraphOne and and Aspen are evolving graph systems, which are representatives of applications that perform analytics over frequently updated datasets. 
GraphOne uses adjacency lists and edge lists to store an input graph while Aspen utilizes compressed purely-functional trees to store a graph, which supports a higher update rate. 
The working sets of these applications change continuously. Their accesses are very complex: 
the first stage builds the graph in memory, exhibiting a random pattern. The second stage runs iterative algorithms where the first iteration does not have locality and thus performs random accesses; the subsequent iterations would enjoy better locality
if it runs on \tool, which dynamically improves the locality during the first iteration. However, updates to the input graph disrupt the locality and hence there can also be many random accesses in the middle of the iterations. 
We used these two graph frameworks to evaluate how well \tool can dynamically adjust the data layout and improve locality.

Third, Metis (MapReduce) and DataFrame represent bulk data processing systems with clear phase-changing behaviors (discussed in \S\ref{sec:motivation}). These workloads are used to evaluate whether \tool can accurately recognize access patterns and switch to the proper data path. DataFrame is additionally used to evaluate compute offloading due to its memory-intensive operations (\S\ref{sec:drill-down}).

Finally, WebService is an interactive web application exhibiting mixed access patterns, from random, pointer-chasing, to sequential accesses.

For \tool to run these applications, we modified 263 lines of code for Metis, 278 lines for Aspen, 219 lines for GraphOne, and 391 lines for Memcached; the additional code was used to declare smart pointers and dereference scopes. It took one developer a few hours to port each program.

\MyPara{Memory setup.}
Each application was run with five local memory configurations: 13\%, 25\%, 50\%, 75\% and 100\%, each representing a specific percentage of an application's working set that can fit into local memory. These configurations were enforced using \codeIn{cgroup}. The first four configurations were employed to evaluate the performance of the three systems when using different amounts of remote memory, while the 100\% (all local memory) configuration was used to assess the runtime overhead of \tool and AIFM, introduced by the barriers (for smart pointer dereferencing), dereference trace recording (for object-level prefetching), and evacuation (for defragmentation), as well as other bookkeeping overheads; see Table~\ref{tab:runtime-overheads} for more details.

\subsection{Throughput\label{sec:eval:tput}}
We first measured the throughput of the applications with varying local memory ratios. Overall, \tool outperforms  Fastswap and AIFM, respectively, by 3.2$\times$ and 1.5$\times$, over the eight real-world applications using remote memory (from 13\% to 75\% local memory). When running locally (100\% local memory),  \tool and AIFM incur an overall overhead of 19.1\% and 14.0\%, respectively, of which 10.2\% and 2.3\% are from the barriers. This section reports the overall performance and runtime overhead.  We show a detailed overhead breakdown in \S\ref{sec:drill-down}.

\begin{figure*}[h]
\vspace{-0.8em}
\begin{center}
\includegraphics[width=0.9\linewidth]{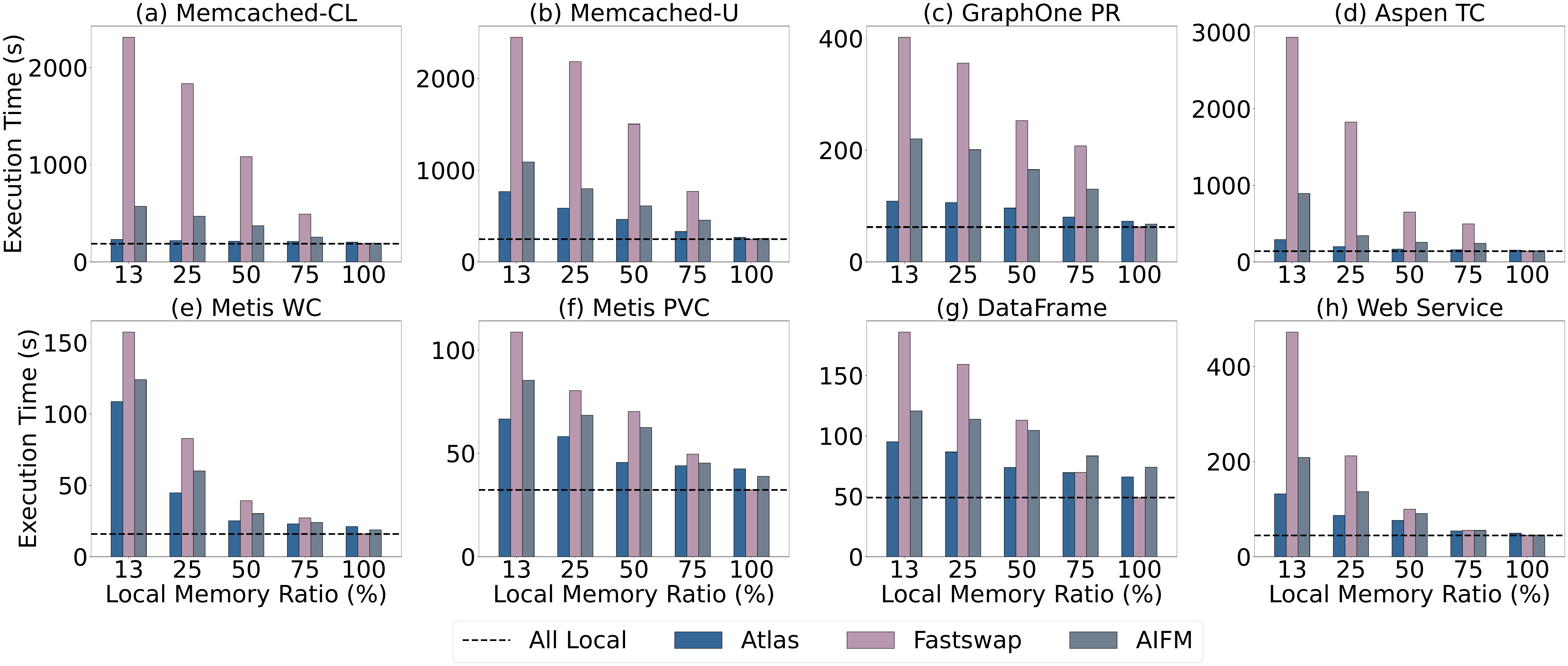}
\vspace{-1.0em}
\caption{Throughput comparison between \tool, Fastswap and AIFM with varying local memory ratios.  "All Local" lines represent the performance of unmodified applications under 100\% local memory.\label{fig:overall-real-apps}}
\vspace{-2em}
\end{center}
\end{figure*}

\MyPara{MCD-CL and MCD-U.}
Both workloads were configured with the same operation ratios, \ie, 87.4\% get and  12.6\% set. As shown in Figure~\ref{fig:overall-real-apps}(a), for a highly-skewed workload like MCD-CL, both \tool and AIFM outperform Fastswap (by 6.4$\times$ and 3.2$\times$, respectively). The performance difference comes primarily from the reduced I/O amplification\textemdash Fastswap fetches 26$\times$ and 30$\times$ more data than \tool and AIFM, respectively, resulting in wasted memory (for storing unused data) and significantly more swaps. 
Under 100\% local memory, \tool and AIFM introduce an overall overhead of 9.0\% and 3.2\%, respectively, compared to Fastswap. The primary source of the overhead is the barriers, taking 6.2\% and 1.5\% of the execution time, respectively. Given that Memcached spends a substantial portion of its execution on communication, the barrier overhead, which is associated with the in-memory processing, is insignificant.

Compared to AIFM, \tool further improves the performance by 1.2$\times$, 1.8$\times$, 2.2$\times$, 2.5$\times$, under the four different memory configurations (75\%, 50\%, 25\%, and 13\%). This improvement stems from a much higher eviction throughput (on average 4.6$\times$ higher) in \tool due to the elimination of object eviction. 
In addition, \tool's concurrent evacuator (\S\ref{sec:memory_management}) improves the temporal locality by segregating hot objects into contiguous pages, leading to an overall of 18\% more accesses that go through the paging path (\S\ref{sec:drill-down}).
This result was achieved when AIFM used 20 eviction threads while \tool only used one single swap-out thread in the paging path. MCD-U performs random accesses with no hot data, hindering opportunities for \tool to  improve locality. Hence, the usefulness of the hybrid data plane is limited. However, \tool still outperforms AIFM by up to 1.4$\times$ due to more efficient eviction, as shown in Figure~\ref{fig:overall-real-apps}(b).

\MyPara{GPR and ATC.}
To execute an evolving graph engine, we divided the input datasets~\cite{Kwak10www} into three batches, which are incrementally fed to the graph engine. For each batch, the graph engine conducts the following three steps: load the updates, update the graph, and execute the analytics. 

As Figure~\ref{fig:overall-real-apps}(c) shows, in the presence of remote memory, \tool outperforms AIFM and Fastswap by an average of 1.8$\times$  and 3.1$\times$, respectively, on GPR. As stated earlier, graph updating and the first iteration of analytics exhibit random access patterns. As such, GPR's throughput under AIFM is 1.7$\times$  higher than under Fastswap. For \tool, when the analytics starts, objects are accessed and reordered by the object fetching in the first few iterations; in the subsequent iterations, pages storing edge objects are switched to using the paging path due to the gradually established locality. As a result, up to 82\% of pages have their PSFs changed during the execution (from object fetching to paging), as demonstrated in Figure~\ref{fig:path-switching}(b). This improves the analytics throughput.

ATC's computation stages and access patterns are both similar to those of GPR. For ATC, the trees storing the graph data are dynamically reorganized by \tool's runtime path, leading to $\thicksim$38\% of pages changing their PSFs (from object fetching to paging). In addition, evacuation improves locality by segregating hot objects from these trees into a few pages, reducing remote memory accesses by 24\%. As demonstrated in Figure~\ref{fig:overall-real-apps}(d), ATC's overall throughput is 2.0$\times$ higher under \tool than under AIFM.

\revise{When running on 100\% local memory, \tool's barrier overheads for both GPR and ATC are modest, 8.2\% and 4.3\%,  due to the high ratio between raw pointer accesses and smart pointer dereferences. Oftentimes, one object dereference (\eg, obtaining a vertex that contains a series of edges) is followed by dozens of raw pointer accesses (\eg, to individual edges). Each dereference scope contains an average of \emph{21} raw pointer accesses. In addition, for ATC, the barrier overhead is further diluted due to its higher computation and memory access costs (from poor spatial locality).}

\MyPara{MWC and MPVC.} 
Figure~\ref{fig:overall-real-apps}(e) and (f) respectively show the performance of MWC and MPVC. As discussed in \S\ref{sec:motivation}, MPVC exhibits a two-phase behavior that can benefit from adaptive path switching, leading to a 1.2$\times$ and 1.4$\times$ improvement, compared with AIFM and Fastswap, respectively. MWC has a similar two-phase behavior with MPVC but exhibits more random accesses in its map phase, resulting in almost no page that can be flipped to  \codeIn{paging}. Compared to AIFM and Fastswap, MWC has 1.2$\times$ and 1.5$\times$ performance improvement, respectively. 

For these two applications, the runtime overhead is relatively high\textemdash 32.0\% (\tool) and 19.2\% (AIFM), under 100\% local memory. These two Metis workloads are both memory-intensive\textemdash they keep scanning data with high parallelism, leading to both high barrier overhead and profiling overhead (\eg, for card profiling and access trace recording, see \S\ref{sec:drill-down}). \tool's barrier overhead reaches up to 16.1\% and 17.4\% for MPVC and MWC, respectively, which are about 4$\times$ higher than that of AIFM.

\MyPara{DF.}
DF is a table-structured in-memory data structure with hundreds of columns and millions of rows, popularized in Pandas~\cite{pandas}. Users can slice data in different ways and run various statistics. As Figure~\ref{fig:overall-real-apps}(g) shows, \tool outperforms AIFM by 1.2$\thicksim$1.4$\times$ in the four remote-memory settings. We ran a client, developed by the AIFM authors, to conduct a series of \emph{Copy} and \emph{Shuffle} operations on \emph{DF}. Similarly to Metis, \emph{DF} demonstrates clear phase-changing behaviors when processing different operations\textemdash a \emph{Copy} operation copies data from a column, exhibiting excellent spatial locality and a clear sequential pattern, while a \emph{Shuffle} operation reorders rows for each column, exhibiting random patterns. \tool achieves superior performance to AIFM and Fastswap, due to its adaptive access path selection.

AIFM suffers a higher runtime overhead (51.4\%) compared to \tool  (34.7\%) despite having a lighter barrier. The reason is that AIFM maintains a remote vector on the memory server for every DataFrame vector to support the eviction of individual objects with varied sizes. During the execution, DataFrame vectors keep getting allocated and resized. As a result, the remote data structure also needs to be frequently resized to maintain a valid mapping from local objects to their remote memory locations. Resizing is a heavy operation as it requires allocating memory and moving all existing objects. Therefore, it becomes a major source of overhead, which can take two-thirds of the runtime overhead under 100\% local memory. On the other hand, under \tool, eviction is handled by the Linux kernel at a fixed page size and there is no need to maintain any remote data structures. 
Note that frequent resizing of data structures was not observed in other applications. For example, for WS, the hash table array is allocated at the start of the application and its size remains fixed throughout the execution.

\MyPara{WS.}
WS is implemented by AIFM's authors to simulate a distributed workload. Each client (thread) sends 32 requests to look up keys in an in-memory hash table and fetches a single 8KB element from an array. This element is then encrypted with Crypto++~\cite{cryptopp} and compressed using Snappy~\cite{snappy} before being sent back to the client. 
We use a 26GB dataset for the evaluation, which is consistent with the dataset used in AIFM \cite{aifm@osdi2020}. Client requests are generated by following a Zipfian distribution. 

As Figure~\ref{fig:overall-real-apps}(h) shows, compared to AIFM, \tool improves WS' performance by an average of 1.3$\times$ with remote memory. This is due to an extremely large number of objects on the LRU list that must be analyzed by AIFM.
AIFM's performance degradation is  primarily due to the compute resource contention between application and evacuation threads (discussed in \S\ref{sec:motivation}), making it hard for evacuation threads to quickly identify and evict cold objects. Consequently, AIFM ends up evicting arbitrary objects to reclaim memory, resulting in data thrashing.
By using paging for eviction, \tool improves the eviction throughput by 5.8$\times$, lifting data eviction efficiency to 5.9 cycles/byte, which is 7.4$\times$ higher than that of AIFM (43.7 cycles/byte).

Atlas and AIFM have relatively low overhead for WS due to the coarse-grained data fetching (8KB element) and the subsequent compute-intensive encryption. As a result, \tool and AIFM  introduce a 10.1\% and 1.9\% runtime overhead under 100\% local memory, respectively.

\subsection{Latency}
\label{sec:eval:lat}

This section evaluates the latency distribution using the two latency-critical applications: WS and MCD-CL.  The 25\% local memory ratio was used in these experiments.

\begin{figure} [h]
    \centering
        \vspace{-0.5em}
    \begin{adjustbox}{max width=\linewidth} 
    \begin{tabular}{cc}

    \includegraphics[scale=.3]{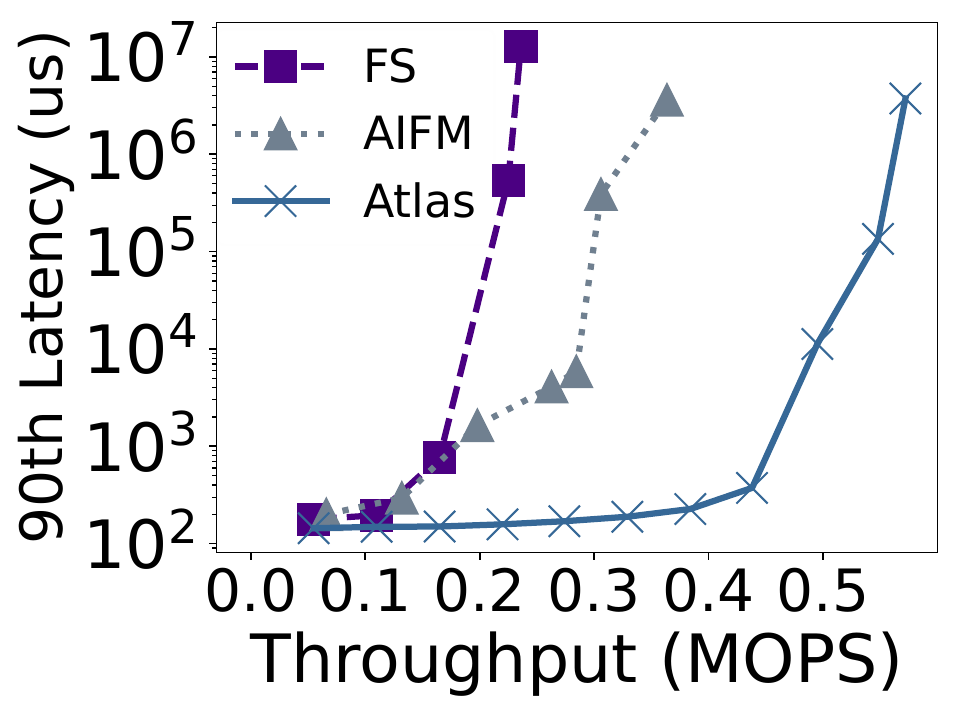} &
    \hspace{-0.7em}
    \includegraphics[scale=.3]{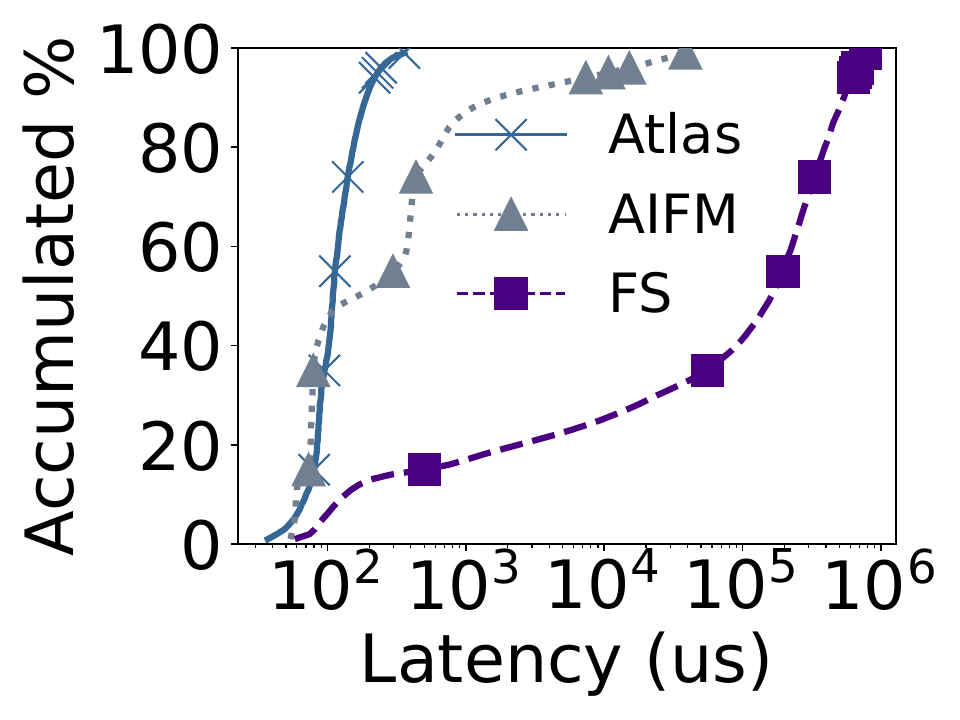}\\[-.2em]
    {\Large (a) 90\upth latency-throughput curve.}& {\Large (b)  Latency CDF.}
    \end{tabular}
    \end{adjustbox}
    \vspace{-0.5em}
    \caption{(a) 90\upth latency as a function of throughput; (b) Latency CDF under 0.23 MOPS offered throughput. FS stands for Fastswap.} \label{fig:latency-web}
    \vspace{-1.5em}
\end{figure}

\MyPara{Web Service (WS).}
Figure~\ref{fig:latency-web}(a) compares the tail latency among the three systems. Fastswap's tail latency rapidly grows due to page thrashing caused by severe access amplification. AIFM reduces amplification so that requests are less blocked by eviction. Despite the reduced amplification, AIFM still has to rank and evict individual key-value pairs, and hence the system saturates at 0.36 MOPS.

\tool fetches individual key-value pairs initially via the runtime path and places those pairs which belong to the same request together on the same page (because these KV pairs are accessed close in time). As the execution progresses, \tool switches to paging that can load multiple key-values pairs at the same time. Meanwhile, page-level eviction continuously offers a much higher eviction throughput so that it never blocks swap-ins. As a result, \tool's tail latency stays low until 0.45 MOPS and can finally reach a peak throughput of 0.57 MOPS. As shown in Figure~\ref{fig:latency-web}(b), the latencies of AIFM and \tool are comparable until the 50th percentile, where the application starts accessing many remote objects leading to increased object management overhead. On the contrary, due to the optimized data layout which enables the efficient use of paging, \tool experiences fewer remote accesses.

\begin{figure} [h]
    \centering
       \vspace{-0.5em}
    \begin{adjustbox}{max width=\linewidth} 
    \begin{tabular}{cc}

    \includegraphics[scale=.3]{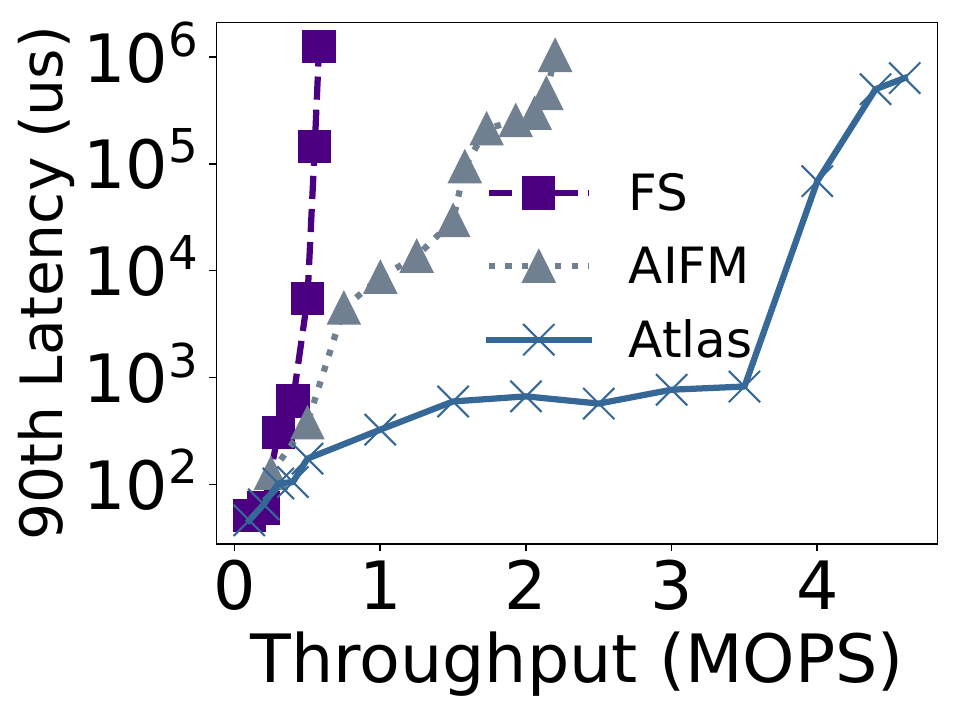} &
    \hspace{-0.7em}
    \includegraphics[scale=.3]{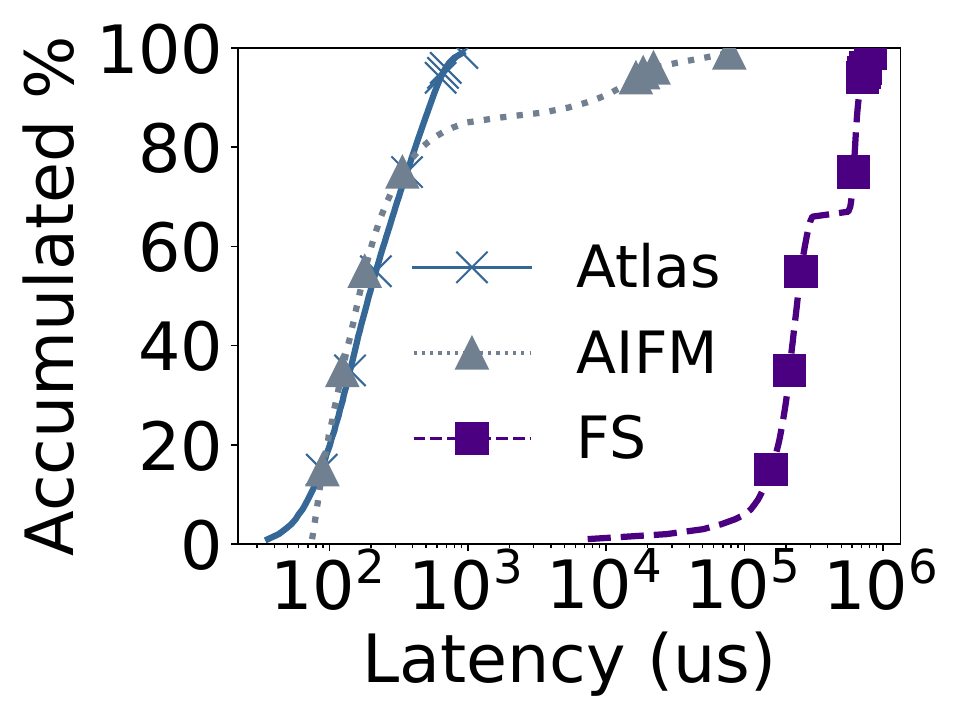}\\[-.2em]
    {\Large (a) 90\upth latency-throughput curve.}& {\Large (b)  Latency CDF.}
    \end{tabular}
    \end{adjustbox}
    \vspace{-1em}
    \caption{(a) 90\upth latency as a function of throughput; (b) Latency CDF under 1 MOPS offered throughput. FS stands for Fastswap.} \label{fig:latency-mcd}
    \vspace{-1em}
\end{figure}

\MyPara{MCD-CL.}
Memcached CacheLib is similar to Web Service as they both access key-value pairs from a hash table. The difference is that every request key in MDC-CL follows a Zipfian distribution, as opposed to accessing key-value pairs always in groups of 32. Figure~\ref{fig:latency-mcd} compares the tail latency among the three systems. It is clear that \tool outperforms the other two systems. In addition to the same reasons explained above, MCD-CL is a skewed workload and hence a substantial portion (40\%) of the improvement comes from the evacuation that groups hot objects in contiguous pages, making these pages amenable to paging.

\subsection{Performance Drill Down \label{sec:drill-down}}
\label{sec:eval:drill-down}

\begin{figure}[h]
\begin{center}
\vspace{-1.3em}
\begin{tabular}{ccc}
\includegraphics[width=0.3\linewidth]{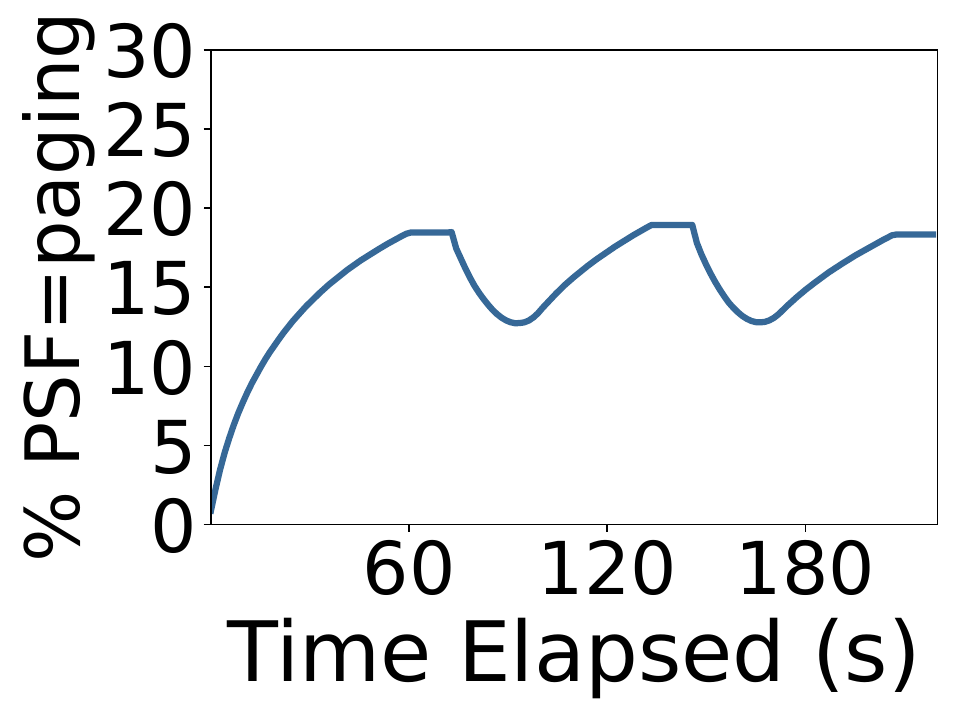} & 
\includegraphics[width=0.3\linewidth]{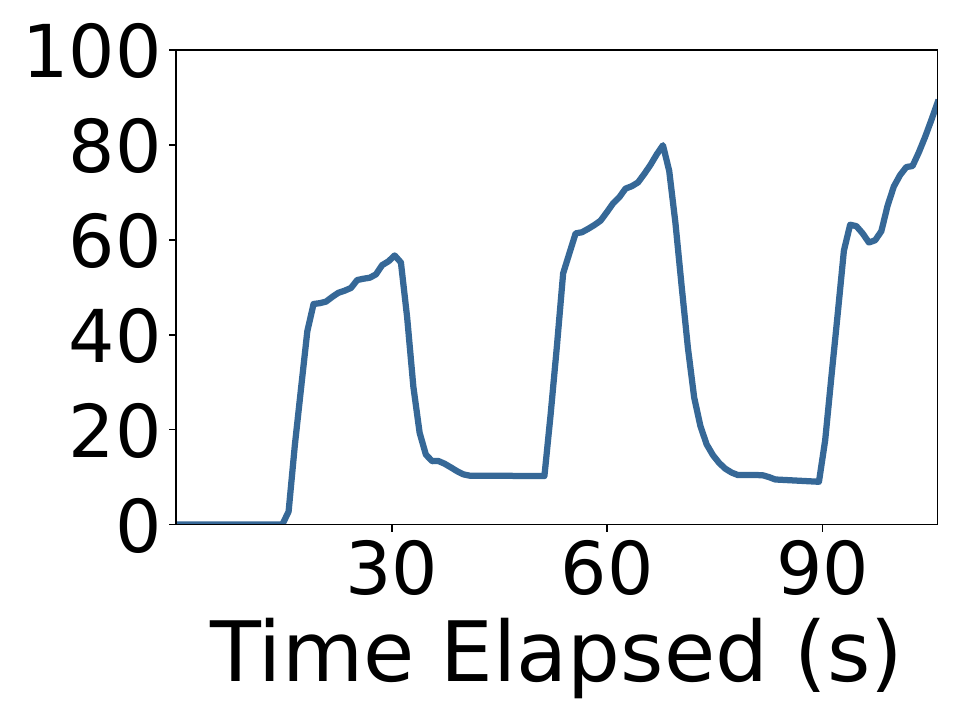} & \includegraphics[width=0.3\linewidth]{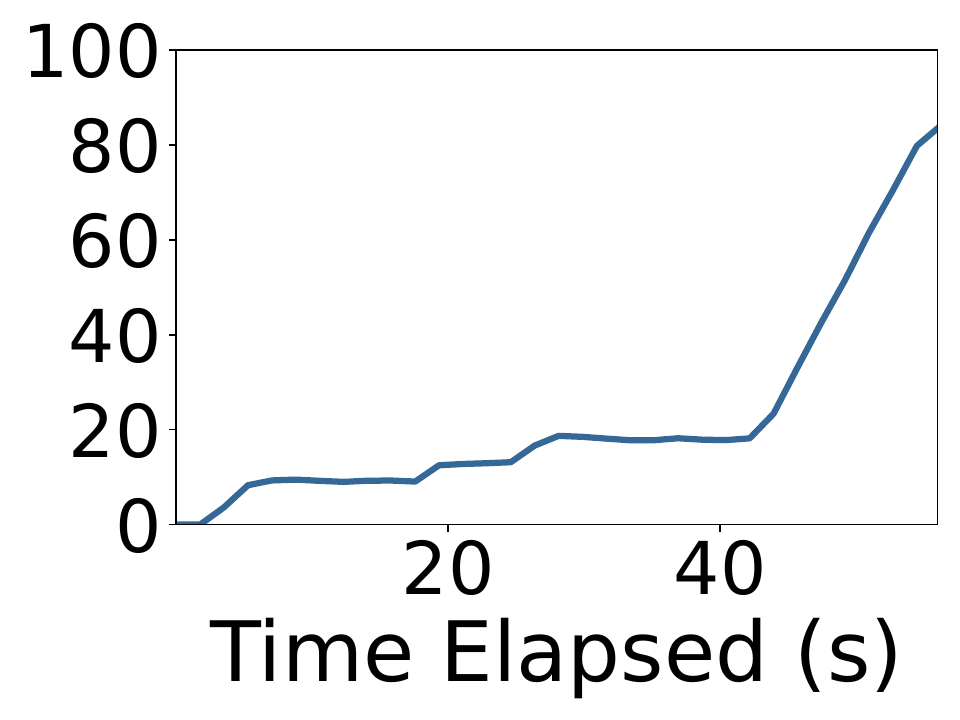} \\
{(a) MCD-CL} & {(b) GraphOne PR} & {(c) MPVC}
\end{tabular}
\vspace{-.5em}
\caption{The percentage of pages with PSF=\camera{\codeIn{paging}} in the memory footprint changes with the elapsed execution time. \label{fig:path-switching}}
\vspace{-2em}
\end{center}
\end{figure}

\MyPara{Adaptive path switching.} To understand the effectiveness of \tool's adaptive path switching, we measured the percentage of the pages whose PSF is \camera{\codeIn{paging} }  during the execution. Figure~\ref{fig:path-switching} demonstrates how this percentage changes during the execution for three applications: 
Memcached CacheLib (MCD-CL), GraphOne Pagerank (GPR) and Metis PageViewCount (MPVC). As Figure~\ref{fig:path-switching}(a) shows, the number of pages that go through the paging path rises and falls over the time due to the \emph{churn} behavior in MCD-CL discussed in \S\ref{sec:setup}. Since the workload is highly skewed, most accesses fall on a small number of hot objects, which stay in local memory and are moved into contiguous pages (with a high CAR) until the hot spot shifts.

As discussed in \S\ref{sec:setup}, the execution of GPR has experienced three batches of updates to the input graph, each of which contains two steps: graph building and analytics. During graph building, applying edge-level updates exhibits random access patterns, which can disrupt locality and leave many pages with a low CAR; these pages would have to go through the object fetching path. However, the subsequent analytics (like PageRank) runs multiple iterations; \tool can quickly improve locality in the first few iterations, making pages turn their PSF to \camera{\codeIn{paging}} in subsequent iterations. This pattern can be clearly seen in Figure~\ref{fig:path-switching}(b). 

MPVC has a clear two-phase behavior (see Figure~\ref{fig:motivation}(a)) which can be accurately recognized by \tool\textemdash the number of pages that go through the paging path increases dramatically as the phase change is detected by \tool (shown in  Figure~\ref{fig:path-switching}(c)).
To understand the individual contributions of object fetching and evacuation to the locality, we disabled the  \codeIn{access} bit tracking and let the evacuator move live objects without guidance. This reduces the overall percentage of pages that go through paging by 4\% on average. 

\begin{figure}[h]
\begin{center}
\includegraphics[width=0.9\linewidth]{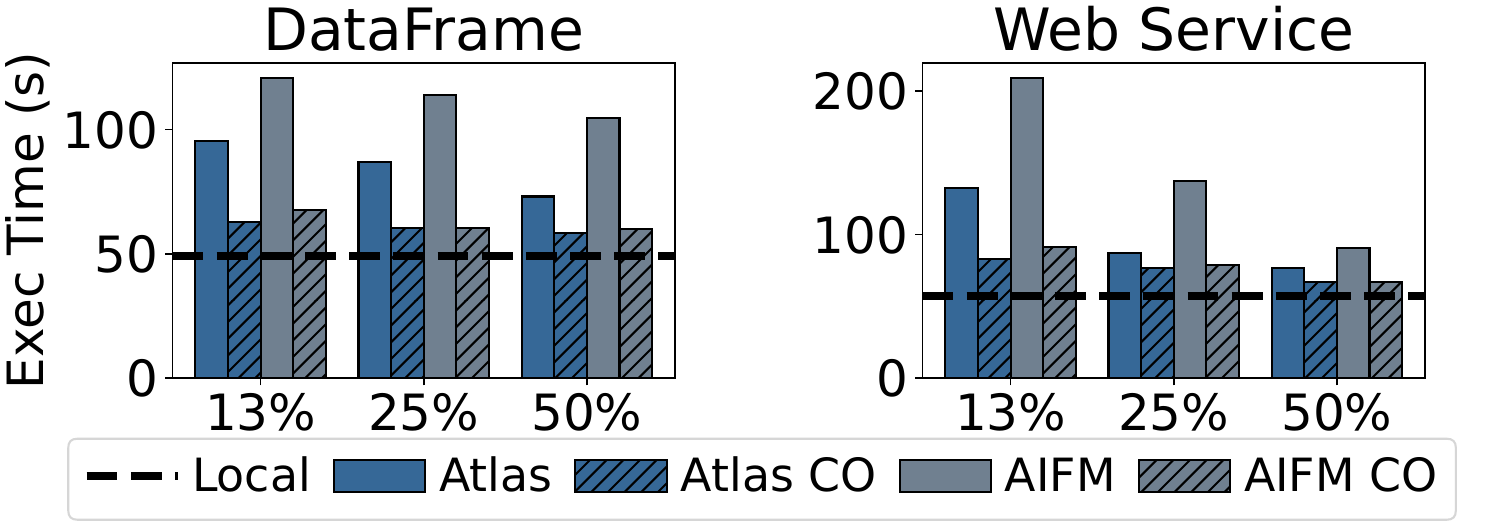}
\vspace{-1.2em}
\caption{Throughput comparisons of DataFrame (DF) and Web Service (WS) when \tool and AIFM enable compute offloading. CO stands for variants with compute offloading. } \label{fig:offload}
\vspace{-2.5em}
\end{center}
\end{figure}

\MyPara{Computation offloading.} 
We compared the offloading performance between \tool and AIFM using DF and WS. 
Figure~\ref{fig:offload} shows the results of \tool and AIFM with and without offloading. 18 cores were reserved on the remote side for both \tool and AIFM, which is consistent with the offloading settings used by AIFM \cite{aifm@osdi2020}. For DF, we offloaded the memory-intensive operations, \ie, \codeIn{Copy} and \codeIn{Shuffle}, to the remote side. For WS, we offloaded the heavyweight array processing (on the 16GB data array).
Compared to the setting where offloading is disabled (Figure~\ref{fig:overall-real-apps} (g) and (h)), the throughputs of \tool and AIFM are both dramatically improved (by up to 1.5$\times$ and 1.9$\times$ for DF, and 1.6$\times$ and 2.3$\times$ for WS, respectively), due to reduced remote accesses and data movement. 
On the other hand, \tool and AIFM achieve comparable performance.
This is because \tool focuses on fetching efficiency; offloading reduces the need for fetching, making \tool's benefit less significant.

\MyPara{Runtime overhead analysis.} To understand the performance penalty introduced by the runtime of \tool and AIFM, we break down and compare the runtime overhead by sources. When running with all local memory, the runtime overhead of \tool and AIFM can be divided into five major components, listed in Table~\ref{tab:runtime-overheads}. Note that the overhead reported here represents the \textbf{worst-case scenario} for \tool when compared against AIFM. When there is remote memory, part of
Atlas's runtime overhead can be eliminated by switching to
the paging path\textemdash dereference trace profiling is not used for paging as its goal is to analyze dereference traces for prefetching objects.
Meanwhile, AIFM incurs more profiling overheads that do not exist under the all local memory setting, such as maintaining the object-level LRU for eviction.

\begin{table}[htbp]
\centering
\begin{adjustbox}{width=\linewidth}

\begin{tabular}{c|c|c}

      \rowcolor[HTML]{EFEFEF} 
      \toprule
      \textbf{Sources of overhead}  & \textbf{Functionality} & \textbf{Affected systems} \\
      \midrule

      {Barrier} & { Correctness guarantee, such as} & { \tool and AIFM} \\ 
      {(Dereferencing)} & {location check \& synchronization} &{}\\
      \midrule
      {Card Profiling} & {Offering data path switching hints.} & {\tool} \\
      \midrule
      {Dereference Trace} & {Offering object-level }  & { \tool and AIFM} \\  
      {Profiling} & {prefetching hints} & {} \\
      \midrule
    {Evacuation} & {Defragmentation}  & { \tool and AIFM}\\
      \midrule
      {Remote Data Structure } & {Managing }  & { AIFM} \\
     {Management} & {object-level eviction} & \\
     
      \bottomrule
             
\end{tabular}
\end{adjustbox}
\caption{\label{tab:runtime-overheads} Major types of runtime overheads, operations involved in each type, and their affected systems. }
\end{table}

\begin{figure}[htbp]
\begin{center}
 \vspace{-1em}
\includegraphics[width=1.\linewidth]{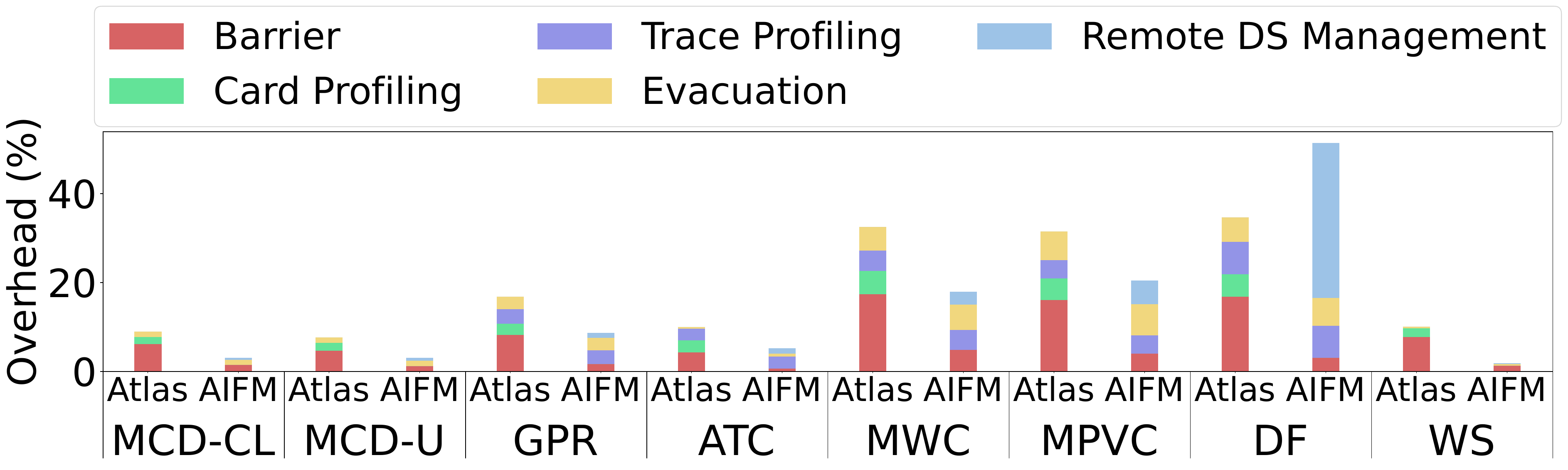}
\vspace{-0.8em}
\caption{\revise{Runtime overhead breakdown: overhead is calculated as the ratio between the extra execution time introduced and the execution time under 100\% local memory.} }\label{fig:revision-overhead-breakdown}
\vspace{-0.8em}
\end{center}
\end{figure}

As shown in Figure~\ref{fig:revision-overhead-breakdown}, compared to Fastswap, the extra tasks in \tool incur a runtime overhead of 7.7-34.7\%, while AIFM's overhead is 1.9-51.4\%. The overall overheads of the two systems are 19.1\% and 14\%, respectively. The primary source of overhead for both systems is the barrier (except for DF with AIFM, for which the reasons are explained in \S\ref{sec:eval:tput}). Specifically, the \tool barrier accounts for half of the total overhead ($\thicksim$10\%), and its cost is 4.4$\times$ of that of AIFM.
Note that this overhead correlates with an application's memory access behavior: the most memory-intensive applications suffer the heaviest barrier overhead (MWC, MPVC, DF).

Although \tool uses a heavier barrier, it underperforms AIFM by \emph{only 4\% under 100\% local memory}. The reason is three-fold: (1) the barrier overhead is effectively amortized across the computation and raw pointer accesses (\S\ref{sec:eval:tput}); (2) AIFM's use of coarse-grained dereference scopes leads to higher synchronization costs than \tool; and (3) there are other operations that also contribute to the runtime overhead. Since the first item has been discussed earlier in this section, here we elaborate on the second and third items.

The barrier conducts two basic tasks, object location checking and synchronization. For location checking, \tool has a much higher overhead than AIFM due to the use of TSX to detect an object's location whereas AIFM checks a bit on each reference. 
However, for synchronization, AIFM's coarse-grained dereference scopes incur a higher cost, which effectively reduces the performance gap between the barriers of the two systems.
After selecting the victim segments, AIFM's evacuator must wait until all application threads exit their dereference scopes to avoid compacting objects being accessed through raw pointers. This design does not work well for big data applications with high object allocation rates, such as MWC, MPVC and Memcached. On the contrary, \tool's fine-grained dereference scope design enables evacuation threads to skip the segments (each aligned to a page in \tool) whose \emph{deref count} is non-zero (indicating they are being used in active dereference scopes) instead of blocking the whole evacuation, leading to significantly reduced synchronization efforts. In fact, \tool's CPU yield rate caused by synchronization is \emph{an order of magnitude lower} than that of AIFM due to our non-blocking design.

Another major source of overhead is the dereference tracing (to provide prefetching hints), accounting for 14\% and 19\% of the total overhead for \tool and AIFM, respectively. 
Among our applications, DF, MWC, MPVC and GPR use array data structures which are amenable to prefetching. As a result, there is a relatively high tracking overhead (accounting for 34\% overhead on average) for both \tool and AIFM.  Other applications such as WS and Memcached use hash maps and small objects as their data structures, which are not as amenable to prefetching as arrays. Hence, for most of their memory accesses, the locations are not tracked and their tracing overhead is much lower.  
Note that with remote memory, the dereference tracing overhead is significantly lower 
under \tool than under AIFM because a large amount of data (\eg, up to 82\% for GPR) goes through the paging path, which utilizes the lightweight page-level prefetcher.

\begin{figure}[t]
\begin{center}
\includegraphics[width=0.7\linewidth]{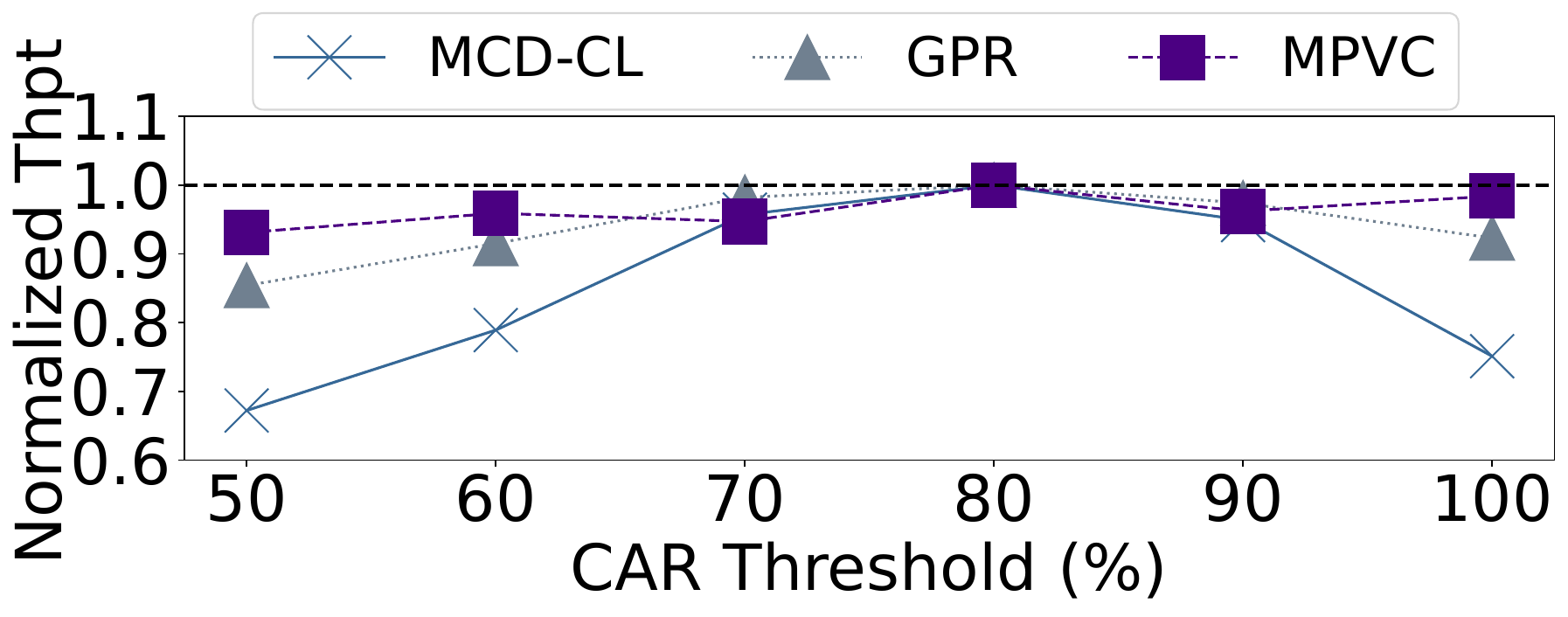}
\vspace{-1.5em}
\caption{Sensitivity of the CAR threshold.  } \label{fig:car-threshold}
\vspace{-2.5em}
\end{center}
\end{figure}

\MyPara{CAR threshold.} Figure~\ref{fig:car-threshold} shows the influence of CAR threshold on the throughput of three applications. Picking the right CAR threshold is a tradeoff between fetching efficiency and resource waste.
We used 80\% as the CAR threshold for flipping PSF in our evaluation. A higher CAR is often too conservative. For example,  in the case of MCD-CL, when the threshold is set to 100\%, we observed that few pages can be flipped to \codeIn{paging}. Therefore, most remote objects still have to be fetched individually instead of fetched in batches with page faults, leading to a 25\% decrease in throughput. On the contrary, a lower CAR may result in premature use of paging, leading to I/O amplification. As shown, the best performance is achieved when the threshold is between 80\% and 90\%. As such, we used the lower bound 80\% based on the observation that the bandwidth of a modern network such as InfiniBand~\cite{infiniband-adapters} is already high and will only become higher in the future, making it possible to transfer (slightly) more data with little overhead.

\begin{figure}[htbp]
\begin{center}
\vspace{-.5em}
\includegraphics[width=0.7\linewidth]{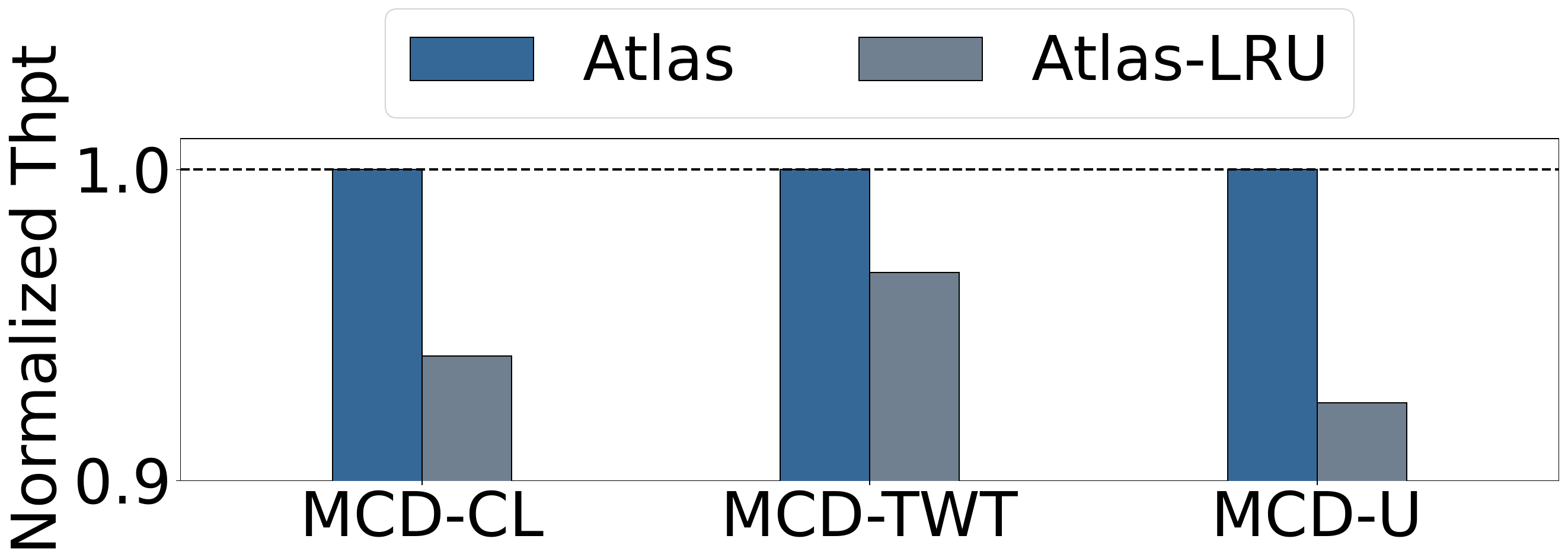}
\vspace{-1.2em}
\caption{\revise{Normalized throughput of Memcached workloads running on \tool and \tool-LRU under 25\% local memory.}} \label{fig:lru-comparison}
\vspace{-2.0em}
\end{center}
\end{figure}

\MyPara{Hotness tracking.} \tool uses an \codeIn{access} bit on each smart pointer to segregate hot and cold objects during evacuation, offering  benefits to workloads that exhibit skewness. 
We evaluated the effectiveness of \tool's \codeIn{access} bit with three  skewed workloads, \ie, highly-skewed (Meta, MCD-CL)~\cite{cachelib}, moderately-skewed (Twitter, MCD-TWT)~\cite{twwiter-trace@nsdi20} and uniformed without skewnewss (MCD-U)~\cite{cooper-ycsb-ds}. We compared \tool with a baseline (Atlas-LRU) equipped with an LRU-like policy from CacheLib~\cite{cachelib@osdi20}, which represents a more accurate approach to identifying hotness.

As shown in Figure~\ref{fig:lru-comparison}, \tool's single-bit design outperforms the LRU-like design by 7.5\%, 3.3\% and 6.0\%, respectively. The LRU-like policy trades compute resources for accuracy by maintaining the logical ordering of objects via a linked list. Each dereference triggers a promotion that moves the object to the head of the LRU list. In order to reduce the overhead, we adopted \emph{flat combining}~\cite{lock-combine} (to reduce thread lock contention) and ignored the dereferences of an object within 10s (to reduce promotion frequency for extremely hot objects)~\cite{cachelib@osdi20}. However, although an LRU-like policy can reduce the frequency of remote access, it incurs a maintenance overhead of up to 9\% due to a huge number of objects.

Of course, the more bits used, the higher accuracy they bring. \tool allows developers to customize the hotness tracking policy with the two reserved bits in each smart pointer (Figure~\ref{fig:atlas-unique-ptr}). For our applications, we did not observe significant performance variations between using one and two access bits\textemdash likely the ability of distinguishing hot and cold objects is not increased much with two access bits.

\section{Related Work} 
\MyPara{Disaggregation.}
Resource disaggregation has become a trending architecture for datacenters to improve resource utilization. Its key idea is to break the server hardware boundary and unstrand idle resources of remote servers by leveraging advanced network hardware~\cite{resource-disaggregation-osdi16, disaggregated_mem@hotnets13}. Existing systems have demonstrated the viability of disaggregated storage~\cite{rack-scale-disaggregated-storage-hotstorage17,reflex-asplos17}, accelerators~\cite{vgpu, ava@hotos19,fractos@eurosys22}, network~\cite{supernic@arxiv}, and memory~\cite{legoos-osdi18, infiniswap-nsdi17}. For a memory-disaggregated system, memory spans across multiple servers. The efficient data path of \tool can speed up the data transfer between servers.

\MyPara{Paging-based far memory.}
A practical way to deliver far memory is to leverage the paging system to access far memory. Google and Meta have reported their successful deployment of such systems in their datacenters~\cite{lagar-cavilla-asplos19, tmo@asplos22}. Many optimizations to the kernel data path have been proposed for improved efficiency, including but not limited to bypassing the block layer~\cite{fastswap-eurosys20, hermit@nsdi23}, prefetching more accurately~\cite{leap-atc20}, and reducing interference~\cite{canvas-nsdi23}. \camera{The design of \tool is orthogonal to the underlying paging systems and can directly benefit from optimizations within these systems.}

\MyPara{Object-based far memory.}
Many runtime libraries offer new primitives for object-granularity far memory management, making them a more efficient alternative for scattered data on far memory. For example, AIFM~\cite{aifm@osdi2020} proposed remote-able data structures, FaRM~\cite{farm-nsdi14} offered key-value interfaces, and Grappa~\cite{sdsm-atc15} builds a software distributed memory. \tool focuses on the cooperative use of its two data paths and benefits directly from existing optimizations.

\MyPara{Emerging hardware.}
Emerging hardware technologies unlock new opportunities for efficient far memory.
Clio~\cite{clio@asplos22}, StRoM~\cite{strom-eurosys20}, and RMC~\cite{rmc@hotnets20} offload functionalities to their customized hardware to reduce network traffic. Finally, CXL~\cite{cxl@atc22,cxl, zhang2023partial, ms-cxl@arxiv, li2023pond} and Project PBerry~\cite{pberry@hotos, kona@asplos21} enable far memory access at the cache-line granularity.
\tool directly benefits from the throughput and latency advancements of new hardware technologies. Besides, for hardware solutions with a fixed access granularity, \tool can improve data locality to improve data transfer efficiency.

\section{Conclusion} 
We present \tool, a hybrid dataplane that enables efficient far memory for bulk data and scattered objects simultaneously. 
\tool outperforms both the state-of-the-art object-based and paging-based far memory systems. 

\section*{Acknowledgement}
We thank the reviewers for their comments and are particularly grateful to our shepherd Malte Schwarzkopf for his feedback. This work is supported by National Key Research and Development Plan of China under grant 2022YFB4500400, National Natural Science Foundation of China under grant 62090024, US National Science Foundation under grants CNS-1763172, CNS-2007737, CNS-2006437, CNS-2106838, CNS-2147909, CNS-2128653, CNS-2301343, CNS-2330831, CNS-2403254, as well as supports from Cisco and Tencent Big Data. 

\appendix
\section{Artifact Appendix}

\subsection{Overview}

Atlas is a kernel-runtime co-designed system to enable a hybrid remote memory data plane. The artifact includes the custom Linux kernel and the runtime library to enable Atlas-managed applications. To run the artifact, two servers with Intel CPUs connected by InfiniBand are required. The server running the application is the CPU server, while the other server providing remote memory is the memory server. Detailed instructions can be found in \tool code repository.

\subsection{Checklist}
\begin{itemize}
  \item {\bf Hardware:} Two servers with Intel CPUs with TSX, connected by InfiniBand
  \item {\bf Software Environment:} Ubuntu 18.04, 20.04 or 22.04, with the specified version of MLNX\_OFED driver and provided Linux kernel described below
  \item {\bf Public Link to Repository:} \url{https://github.com/wangchenxi7/Atlas}
  \item {\bf Code License:} MIT License
\end{itemize}

\subsection{Building the Linux Kernel}

\begin{mdframed}[style=MyFrame]
\codeIn{\#\# all operations are performed on both servers unless specified}\\
\codeIn{cd linux-5.14-rc5}\\
\codeIn{cp config .config}\\
\codeIn{sudo apt install -y build-essential bc python2 bison flex libelf-dev libssl-dev libncurses-dev libncurses5-dev libncursesw5-dev}\\
\codeIn{./build\_kernel.sh build}\\
\codeIn{./build\_kernel.sh install}\\
\codeIn{./build\_kernel.sh headers-install}\\
\codeIn{\#\# edit GRUB\_DEFAULT="Advanced options for Ubuntu>Ubuntu, with Linux 5.14.0-rc5+", or whatever the new kernel version code is}\\
\codeIn{\#\# edit GRUB\_CMDLINE\_LINUX="nokaslr transparent\_hugepage=never processor.max\_cstate=0 intel\_idle.max\_cstate=0 tsx=on tsx\_async\_abort=off mitigations=off"}\\
\codeIn{sudo vim /etc/default/grub}\\
\codeIn{sudo update-grub}\\
\codeIn{sudo reboot}
\end{mdframed}

\subsection{Setting up InfiniBand Connection}

\begin{mdframed}[style=MyFrame]
\codeIn{\#\# use Ubuntu 18.04 as an example below}\\
\codeIn{wget https://content.mellanox.com/ofed/\\MLNX\_OFED-5.5-1.0.3.2/MLNX\_OFED\_LINUX-5.5-\\1.0.3.2-ubuntu18.04-x86\_64.tgz}\\
\codeIn{tar xzf MLNX\_OFED\_LINUX-5.5-1.0.3.2-\\ubuntu18.04-x86\_64.tgz}\\
\codeIn{cd MLNX\_OFED\_LINUX-5.5-1.0.3.2-\\ubuntu18.04-x86\_64}\\
\codeIn{sudo apt install -y bzip2}\\
\codeIn{sudo ./mlnxofedinstall --add-kernel-support}\\
\codeIn{sudo /etc/init.d/openibd restart}\\
\codeIn{sudo update-rc.d opensmd remove -f}\\
\codeIn{sudo sed "s/\# Default-Start: null/\# Default-Start: 2 3 4 5/g" /etc/init.d/opensmd -i}\\
\codeIn{sudo systemctl enable opensmd}\\
\codeIn{sudo service opensmd start}\\
\codeIn{\#\# assign IPs to InfiniBand interfaces on both servers}\\
\codeIn{sudo nmtui}
\end{mdframed}

\subsection{Building Atlas Runtime}

\begin{mdframed}[style=MyFrame]
\codeIn{\#\# use gcc-9}\\
\codeIn{cd atlas-runtime/third\_party}\\
\codeIn{git clone --depth 1 -b 54eaed1d8b56b1aa528be3bdd1877e59c56fa90c https://github.com/jemalloc/jemalloc.git}\\
\codeIn{cd ../bks\_module/remoteswap}\\
\codeIn{\#\# on memory server}\\
\codeIn{cd server \&\& make}\\
\codeIn{\#\# on CPU server}\\
\codeIn{cd client \&\& make}\\
\codeIn{cd ../../bks\_drv \&\& make}\\
\codeIn{cd ../.. \&\& mkdir build \&\& cd build}\\
\codeIn{cmake .. \&\& make -j}
\end{mdframed}

\subsection{Running Atlas Applications}

\begin{mdframed}[style=MyFrame]
\codeIn{cd atlas-runtime/bks\_module/remoteswap}\\
\codeIn{\#\# on memory server}\\
\codeIn{cd server}\\
\codeIn{\#\#./rswap-server <memory server IB ip> <memory server IB port> <memory pool size in GBs> <CPU server core count> e.g.,}\\
\codeIn{./rswap-server 172.16.16.1 9999 48 96}\\
\codeIn{\#\# on CPU server}\\
\codeIn{cd client}\\
\codeIn{\#\# edit `mem\_server\_ip`, `mem\_server\_port` and `SWAP\_PARTITION\_SIZE\_GB` to be consistent with memory server parameters}\\
\codeIn{vim manage\_rswap\_client.sh}\\
\codeIn{bash manage\_rswap\_client.sh install}\\
\codeIn{\#\# run a test}\\
\codeIn{cd atlas-runtime/build/tests/\\runtime/unique\_ptr}\\
\codeIn{bash test.sh ./unique\_ptr\_test}
\end{mdframed}

\bibliographystyle{abbrv}

\end{document}